\documentclass[a4paper,10pt]{article}
\usepackage{jcappub} 
\usepackage{lineno}
\usepackage{tabularx}

\arxivnumber{...} 
\title{\boldmath Bound Dark Energy: Particle Physics model in alignment with recent DESI cosmological measurements}

\author{Axel de la Macorra, Jose Agustin Lozano Torres}
\affiliation{Instituto de Física, Universidad Nacional Autónoma de México\\
Circuito de la investigación científica 04510, Ciudad de México, México}

\emailAdd{lozanotorres@estudiantes.fisica.unam.mx}

\abstract{We present observational constraints on the Bound Dark Energy Cold Dark Matter (BDE-CDM) model using DESI DR2 baryon acoustic oscillation measurements combined with Planck CMB data and Type Ia supernovae compilations (PantheonPlus, Union3, DESY5). In BDE-CDM, dark energy originates from the lightest meson field within a supersymmetric SU(3) dark gauge group with $N_f= 6$ flavors, governed by an inverse power-law potential $V(\phi)= \Lambda_{c}^{4+2/3} \phi^{-2/3}$. Unlike $\Lambda$CDM and $w_0w_a$CDM, the dark energy sector contains no free parameters—the condensation scale $\Lambda_c$ and transition epoch $a_c$ are determined by gauge coupling unification constraints. The equation of state evolves from relativistic behavior ($w = 1/3$) before condensation through a kinetic-dominated stiff phase ($w\simeq 1$), approaching $w_0 = -0.9298 \pm 0.0003$ at present, with $w> -1$ maintained throughout cosmic history, avoiding phantom-regime instabilities. We obtain $\Lambda_{c} = 43.93 \pm 0.13$ eV and $a_c = (2.489 \pm 0.007) \times 10^{-6}$, consistent with theoretical predictions. The $w_0$-$w_a$ confidence contours are approximately 10,000 times smaller than those of $w_0w_a$CDM while achieving comparable fits, and remain stable across different supernova datasets. Statistical analysis yields $\Delta$DIC = -6.77 and $\Delta$AIC = -8.97 relative to $\Lambda$CDM for BAO+DESY5, constituting strong evidence favoring BDE-CDM model. The model predicts distinctive signatures including $25\%$ enhancement in the matter power spectrum at $k \approx 4.3\text{ } h\text{ } \text{Mpc}^{-1}$. These results establish BDE-CDM as a theoretically motivated framework that successfully addresses the DESI-observed preference for dynamical dark energy while connecting particle physics with cosmological observations. }
\begin{document}
\maketitle
\flushbottom

\section{Introduction}
\label{sec:intro}

The cosmological constant $\Lambda$, identified as the driver of cosmic acceleration through independent observations \cite{Riess_1998,Schmidt_1998,Perlmutter_1999}, has been widely regarded as the candidate for dark energy, accounting for $68\%$ of the Universe's total energy density at present time. For more than two decades, the standard cosmological model $\Lambda$CDM, which incorporates a cosmological constant $\Lambda$ and cold dark matter (CDM), has achieved strong agreement with a wide range of cosmological observations \cite{Bennett_2013, refId0}. However, the Dark Energy Spectroscopic Instrument (DESI) precise measurements of baryon acoustic oscillations (BAO) combined with \textit{Planck} CMB data and three different samples of Type Ia SNe distance moduli measurements (Pantheon-Plus \cite{Scolnic_2022}, Union3 \cite{Rubin_2025} and DESY5 \cite{Abbott_2024}), reveal a preference --- though not yet firm evidence --- for dynamical dark energy (DDE), i.e. a time-varying dark energy equation of state (DE-EoS) parameter $w(z)$, at a statistical level from $2.8\sigma$ to $4.2\sigma$ away from the cosmological constant value ($w \equiv -1$), depending on the combination of SNe data used \cite{Adame_2025}. To assess the dynamics of DE, DESI collaboration uses a simple parameterization of the evolution of DE, namely $w_0w_a$CDM, where $w_0$ and $w_a\equiv -dw/da|_{a_{0}}$ (with $a_{0}\equiv1$) are the EoS parameter and its first derivative at present time \cite{PhysRevLett.90.091301}. This simple ansatz allows DESI to discard a cosmological constant as dark energy at more than 95\% level opening a fundamental question regarding the nature of dark energy.

Here, we analyzed our cosmological BDE-CDM model, where the dark energy component in $\Lambda$CDM or $w_0w_a$CDM models is replaced by our Bound Dark Energy (BDE) model \cite{PhysRevD.72.043508, PhysRevLett.121.161303, PhysRevD.99.103504} supplemented by cold dark matter (CDM) component in the dark sector. In our BDE-CDM model, the dark energy component corresponds to the lightest meson field denoted by $\phi$, residing within a supersymmetric SU(3) dark gauge group (DGG) with $N_f=6$ flavors  (c.f. \cite{PhysRevLett.121.161303,PhysRevD.72.043508, PhysRevD.99.103504}). Interesting, our BDE model is described by the same mathematical formalism as in the well established Standard Model (SM) of particle physics, namely in terms of gauge groups  which encompasses the gauge interactions SU(3)$\times$SU(2)$\times$U(1) corresponding to the strong, weak and electromagnetic forces. Interesting, our BDE-CDM was proposed in 2018 \cite{PhysRevLett.121.161303, PhysRevD.99.103504}, previous to DESI observations and results \cite{tr6ykpc6,Adame_2025}, and it is only recently (2024-2025) that the precise DESI measurements allows to determine the dynamics of DE ruling out the cosmological constant at more than 95\% confidence level \cite{tr6ykpc6,Adame_2025}. 

The dark energy sector in our BDE-CDM model contains no free parameters, while $\Lambda$CDM model includes one additional free parameter, the well-known cosmological constant $\Lambda$ (with energy density $\rho_{\Lambda}=\Lambda^{4}$) and $w_0w_a$CDM model which has three free parameters in the dark energy sector namely $\rho_{\mathrm{DE}}(a_{0}), w_0$, and $w_a$. 
Analyzing the recent BAO measurements observed by DESI \cite{tr6ykpc6}, in combination with cosmic microwave background (CMB) data \cite{refId0} and the three compilation of Type Ia supernova distance measurements --PantheonPlus \cite{Scolnic_2022}, Union3 \cite{Rubin_2025} and DESY5 \cite{Abbott_2024}—  we obtain cosmological constraints on the base and derived key parameters according to our BDE-CDM model and delineate the differences with the resulting values and constraints derived from $\Lambda$CDM and the $w_0w_a$CDM models, as deeply analyzed by the DESI collaboration \cite{Adame_2025, tr6ykpc6}.

\section{Theoretical framework of BDE-CDM model}
\label{Theoretical Background}
The BDE-CDM model proposes that source of dark energy corresponds to the lightest meson field denoted by $\phi$ emerging dynamically a due to a strong gauge coupling constant within a dark gauge group as the universe expands and cools down, driving the late-time acceleration of the universe observed recently by DESI \cite{Adame_2025}.
This emergence is driven due to non-perturbative gauge dynamics of an $SU(N_c=3)$ supersymmetric dark gauge group with $N_f=6$ flavours,  at a strong  gauge coupling regime occurring at low energies \cite{PhysRevLett.121.161303}. 
As a result, a phase transition occurs at the condensation scale $\Lambda_c$ with a scale factor denoted by $a_{c}$ and where the original fundamental particles bind together forming neutral composite states e.g. dark mesons. This process is similar as in the strong QCD interactions of the Standard Model (SM), where neutral composite particles (baryons and mesons) are generated from underlying quark constituents. The evolution of the lightest meson field, denoted by $\phi$, is governed by a non-perturbative inverse-power-law (IPL) scalar potential $V(\phi)$, generated within a strong gauge coupling regime in a supersymmetric gauge group \cite{BURGESS1997181,AFFLECK1984493}. The resulting scalar potential $V(\phi)$ is described by \cite{PhysRevD.72.043508,PhysRevLett.121.161303}:
	\begin{equation}\label{eq1}
		V(\phi)=\Lambda_{c}^{4+n}\phi^{-n}=\Lambda_{c}^{4+2/3}\phi^{-2/3},
	\end{equation}
	\noindent 
    for a  $SU(N_c=3)$ supersymmetric gauge group with particle content $N_f=6$ giving an inverse power law exponential potential  $n=2(N_c-N_f)/(N_c+N_f)=2/3$ for our BDE model \cite{2003JHEP...01..033D, PhysRevLett.121.161303}. The condensation energy scale $\Lambda_c$ determines the scale where the $SU(3)$ gauge coupling constant becomes strong and neutral composite particles are dynamically formed. In our model it is given by 
    \begin{equation}\label{eq2}
        \Lambda_c=\Lambda_{\mathrm{gut}} e^{-8\pi^2/(b_0 g _{\mathrm{gut}}^2)} = 34^{+16}_{-11}\mathrm{eV},
    \end{equation}
    with $b_0=3N_c-N_f=3$,  with
	$g^2_{gut}= 4\pi/(25.83\pm 0.16)$  and  $\Lambda_{\mathrm{gut}}=(1.05\pm 0.07)\times 10^{16}$ GeV, the strength of the coupling constant and the unification scale, respectively, as given by \cite{Bourilkov_2015,PhysRevD.110.030001}. 
    This framework parallels the established SM of particle physics, encapsulated by $SU_{\text{QCD}}(N_{c}=3) \times SU(N_{c}=2)_{L}\times U_{Y}(N_{c}=1) $, which incorporates three generations of particles and dictates the strong, weak and electromagnetic interactions. We emphasize that the choice of gauge groups and particle content in the SM specifies the theoretical constructs but does not arise from any underlying fundamental theory, setting an equivalent principle between BDE and the SM particles.\newline
    The BDE energy density before the condensation scale ($a \leq a_{c} $) consists of relativistic DGG particles, implying the addition of an extra amount of radiation of $N_{\mathrm{ext}}=0.945$ \cite{PhysRevD.99.103504}, remaining consistent within the observational limits \cite{Mattias_Blennow_2012,MANGANO2011296,Jan_Hamann_2011}. For $a<a_c$ the energy density of the DGG is expressed as: $\rho_\mathrm{DGG}(a)=\rho_\mathrm{DGG}(a_c)(a_c/a)^{4} = 3\Lambda_c^4  (a_c/a)^{4}$, with $\rho_\textrm{DG}(a_c) =3\Lambda_c^4$. Therefore we have $\rho_\mathrm{DGG}(a_c)/\rho_r(a_c)=3\Lambda_c^3/(\rho_{r0}a_c^{-4})=3(a_c\Lambda_c)^4/\rho_{r0}$, where $\rho_{r}$ accounts for relativistic particles with the present-day radiation energy density $\rho_{r0}=(\pi^{2}/15)g_{r}T^{4}_{0}$ with $T_{0}=2.7255 \pm 0.0006 \textbf{ }\mathrm{K}$ \cite{Fixsen_2009}, where $g_{r}=2+\frac{7}{8}(2)N_{\nu}(T_{\nu}/T_{\gamma})^{4}=3.383$ and $N_{\nu}=3.046$ accounts for neutrino decoupling effects \cite{MANGANO20028}. Solving for $a_c\Lambda_c$, we get the constraint equation \cite{PhysRevD.99.103504,PhysRevLett.121.161303}:
	\begin{align}\label{eq:bde_acLc_theory}
		\frac{a_c\Lambda_c}{\textrm{eV}} & =  \Big( \frac{\rho_{r0}}{3\mathrm{eV^{4}}}\frac{g^{\mathrm{GUT}}_{\mathrm{DGG}}}{g_{r}} \Big)^{\frac{1}{4}}\Big(\frac{4}{11}\frac{g^{\nu\mathrm{dec}}_{\mathrm{SM}}}{g^{\mathrm{GUT}}_{\mathrm{SM}}}\Big)^{\frac{1}{3}}= 1.0939 \times 10^{-4},
	\end{align}
	\noindent 
	where $g^{\mathrm{GUT}}_{\mathrm{DGG}}=97.5$, $g^{\mathrm{GUT}}_{\mathrm{SM}}=228.75$, $g^{\nu\mathrm{dec}}_{\mathrm{SM}}=10.75$ are the relativistic degrees of freedom (d.o.f) of the DGG and SM gauge groups at the unification scale, and neutrino decoupling, respectively \cite{PhysRevD.99.103504}. Eq.(\ref{eq:bde_acLc_theory})  is a meaningful prediction relating the two characteristic quantities of our BDE model, namely $a_c$ and $\Lambda_c$ given in Eq.(\ref{eq2}). Using Eq.(\ref{eq2}) and Eq.(\ref{eq:bde_acLc_theory}) we obtain the transition scale value $a_{c}=(3.22^{+1.54}_{-1.03})\times 10^{-6}$, hence these two equations determine the values of $a_{c}$ and $\Lambda_{c}$, and therefore, the dark energy sector in our BDE model has no free parameters. 
    
    At the phase transition at $\Lambda_{c}$ with scale factor $a_{c}$, the DGG relativistic constituents bind together forming neutral composite states, similar as in QCD, i.e. strong force in the standard model of particle physics. We assume that all of the energy stored in the Dark Gauge Group is  transferred to the lightest particle, corresponding to the dark energy meson filed, named Bound Dark Energy (BDE) \cite{PhysRevD.99.103504, PhysRevLett.121.161303}. Once the condensation occurs, the evolution of dark energy particle BDE is described by a scalar field with energy density and pressure given by $\rho_{\mathrm{BDE}} = \dot{\phi}^{2}/2 + V(\phi), \text{ } P_{\mathrm{BDE}} = \dot{\phi}^{2}/2-V(\phi)$ \cite{PhysRevLett.121.161303}. The dark energy EoS of our BDE  model, $w_{\mathrm{BDE}}=P_{\mathrm{BDE}}/\rho_{\mathrm{BDE}}$, is time-dependent and influenced by the interplay between the kinetic $\dot{\phi}^{2}/2 $ and potential $V(\phi)$ terms.
    The dynamics of BDE-CDM model is set in a Friedmann-Lemaître-Robertson-Walker (FLRW) metric, is described by the equation $\ddot{\phi} + 3H\dot{\phi} + \mathrm{d}V/\mathrm{d}\phi = 0$, with the Hubble parameter defined as $H \equiv \dot{a}/{a} = \sqrt{8\pi G \rho_{\mathrm{Tot}}/3}$. The total energy density of the universe can be summarized as $\rho_{\mathrm{Tot}}(a)= \rho_{m0} a^{-3} + \rho_{r0} a^{-4} + \rho_{\mathrm{BDE}}$. Given the choice of the $SU(N_c=3)$ dark gauge group with $N_{f}=6$ and assuming gauge coupling unification between the BDE and the SM gauge groups, the initial conditions for BDE are given and therefore our BDE model contains no free parameters in the DE sector. 
\section{Datasets and Methodology}
\label{methodology and datasets}
The cosmological datasets utilized to investigate the observational implications and derive parameter constraints for the BDE-CDM, $\Lambda$CDM, and $w_0w_a$CDM models originate from several cutting-edge astrophysical surveys. These are described as follows:
\begin{itemize}
    \item \textit{Planck}: Measurements of the Planck CMB spectrum of temperature anisotropies, $\mathcal{C}^{TT}_{\ell}$, at large scales ($2\leq\ell \leq 30$) obtained from the Planck 2018 legacy survey data released (PR3) \texttt{commander} likelihood \cite{refId0, refId0_1}, measurements of the power spectra of temperature and polarization anisotropies, $\mathcal{C}^{TT}_{\ell}$, $\mathcal{C}^{TE}_{\ell}$, and $\mathcal{C}^{EE}_{\ell}$, at small scales ($\ell > 30$), from the Planck (PR3) \texttt{plik} likelihood \cite{refId0, refId0_1} and measurements of the spectrum of E-mode polarization, $\mathcal{C}^{EE}_{\ell}$, at large scales ($2\leq\ell \leq 30$), from the Planck (PR3) \texttt{SimAll} likelihood \cite{refId0, refId0_1}.

    \item \textit{DESI}: Baryon acoustic oscillations (BAO) measurements extracted from observations of galaxies $\&$ quasars and Lyman-$\alpha$ tracers in the redshift range $0.1<z<4.2$ from the second data release (DR2) using the Dark Energy Spectroscopic Instrument (DESI) \cite{tr6ykpc6}.

    \item \textit{DESY5}: Distance moduli measurements of 1635 Type Ia SNe covering the redshift range of $0.10 < z < 1.13$ that have been collected during the fifth year of the Dark Energy Survey program \cite{Abbott_2024}, along with the  194 low-redshift SNe in the redshift range of $0.025 < z < 0.1$, where 8 come from the Carnegie Supernova Project \cite{Krisciunas_2017}, 68 from the Center for Astrophysics \cite{Hicken_2009,Hicken_2012}, and 118 from the Pan-STARRS Supernova Survey \cite{10.1093/mnras/stx3136}. These 194 low-redshift SNe are in common with the PantheonPlus sample \cite{Scolnic_2022, Brout_2022}.

    \item  \textit{Union3}: The \textit{Union3} compilation \citep{Rubin_2025} is an updated dataset of 2,087 Type Ia supernovae assembled from 24 surveys, spanning redshifts from $z \approx 0.01$ to $z > 1.7$. Light curves are uniformly fit using the SALT3 model, and distance moduli are derived within the UNITY1.5 Bayesian framework, which jointly handles standardization, selection effects, and systematic uncertainties. Data products are publicly available via Zenodo.\footnote{\url{https://doi.org/10.5281/zenodo.14090777}}

    \item \textit{PantheonPlus}: The \textit{PantheonPlus} dataset comprises 1,701 light curves from 1,550 spectroscopically confirmed Type Ia supernovae (SNe Ia). As the direct successor to the original Pantheon analysis, this compilation integrates data from 18 distinct surveys, including the Foundation Supernova Survey, the Dark Energy Survey (DES), and the Lick Observatory Supernova Search. The sample covers a broad redshift range from $z=0$ to $z=2.3$, with a significant increase in the number of low-redshift objects ($z < 0.01$).
\end{itemize}
For parameter estimation and constraint derivation, we implemented an advanced Markov Chain Monte Carlo sampling approach, utilizing the cosmological Boltzmann codes \texttt{CAMB} \cite{Lewis_2000, Lewis_2002} and \texttt{CosmoMC} \cite{PhysRevD.66.103511}, with appropriate modifications to accommodate our BDE-CDM model. The Metropolis-Hastings MCMC algorithm was deployed to execute four parallel independent chains for each combination of dataset and cosmological model, initialized with proposal covariance matrices obtained from preliminary runs. Chain execution continued until satisfying the standard convergence criterion of Gelman-Rubin statistic $R - 1 < 0.01$. Throughout this analysis, we adopt a spatially flat universe ($\Omega_k = 0$) for all cosmological models under consideration. In particular, DESI collaboration \cite{Adame_2025, tr6ykpc6} compares two primary cosmological models: flat $\Lambda$CDM and flat $w_0w_a$CDM. The flat $\Lambda$CDM model is specified by six free parameters: the physical baryon density $\omega_b$ and cold dark matter density $\omega_c$; the angular scale of the sound horizon $\theta_{\rm MC}$; the optical depth to reionization $\tau$; the primordial scalar perturbation amplitude $A_s$; and the scalar spectral index $n_s$. The flat $w_0w_a$CDM model extends this by incorporating two additional free parameters: the present-day dark energy equation of state $w_0$, and $w_a$, which characterizes the temporal evolution of dark energy, yielding eight total parameters. On the other hand, in the BDE-CDM model, as detailed in Section \ref{Theoretical Background}, the DE sector is governed by a specific set of equations and constraints that predicts the amount of dark energy today excluding free parameters in the DE sector. These include the scalar potential defined in Eq.~(\ref{eq1}), the $\Lambda_{c}$ constraint from Eq.~(\ref{eq2}), and the $a_c\Lambda_c$ constraint given by Eq.~(\ref{eq:bde_acLc_theory}). Using the predicted central value and uncertainties for $\Lambda_{c}$ from Eq.~(\ref{eq2}), the transition scale factor $a_{c}$ is determined as $a_{c} = 1.0939 \times 10^{-4}/\Lambda_{c} = (3.22^{+1.54}_{-1.03}) \times 10^{-6}$. 

A fundamental distinction between BDE-CDM and the standard $\Lambda$CDM model lies in the nature of the dark energy. While $\Lambda$ is an unknown free parameter in $\Lambda$CDM, $\Lambda_{c}$ is determined by gauge coupling unification (Eq.~\ref{eq2}) and is therefore not a free parameter in the traditional sense. Consequently, in our MCMC exploration, we substitute $H_0$ (or equivalently $\theta_{\rm MC}$) with $\Lambda_c$ as a primary parameter.  However, upon deeper examination, the essential distinction of the BDE-CDM model is that $\Lambda_c$ does not work as a conventional free parameter. Within the theoretical framework, the value of $\Lambda_c$ is predicted by the condensation scale equation (see Eq.(\ref{eq2})), with uncertainties propagating from measurements of the Grand Unified Theory (GUT) scale parameters $\Lambda_{\rm GUT}$ and $g^2_{\rm GUT}$ \cite{Bourilkov_2015}. To properly propagate these theoretical uncertainties into cosmological predictions, we vary $\Lambda_c$ within its predicted range rather than fitting it as an unknown constant. This implies that the BDE-CDM model, in principle, possesses no free dark energy parameters. To account for the prediction uncertainties, we impose a flat prior $\mathcal{U}[23, 50]$ for $\Lambda_c$ in our analysis. Therefore, the resulting parameter set for the BDE analysis is $ \mathcal{P}_{\rm BDE} = \{\omega_b, \omega_c, \tau, n_s, A_s\}$ since we do not take into account $\Lambda_{c}$ as a free parameter in a strictly way and therefore, our $ \mathcal{P}_{\rm BDE}$ set contains five free parameters, having one and three free parameters less than $\Lambda$CDM and $w_0 w_a$CDM models, respectively. 


\begin{table}
\hspace{-0.9cm}
    \begin{tabular}{|c|c|c|c|c|c|c|c|c|}
    \hline
       \textbf{Parameter}& $w_0$  & $w_a$  & $\omega_{b}$ & $\omega_{c}$ & $100\theta_{\text{MC}}$ & $\tau$ & $\mathrm{ln}(10^{10}A_{s})$ & $n_{s}$   \\ \hline
       \textbf{Prior}     & [-3, 1] & [-3, 2] & [0.005, 0.1] & [0.001, 0.99] & [0.5, 10] & [0.01, 0.8] & [1.61, 3.91]  & [0.8, 1.2] \\ 
    \hline
    \end{tabular}
    \caption{Cosmological parameters and priors ranges.}
    \label{tab:priors}
\end{table}
For the remaining parameters, we employ identical priors to those adopted by the DESI collaboration as shown in Table~\ref{tab:priors} (see Table 2 in \cite{Adame_2025}). To visualize confidence intervals and likelihood distributions of cosmological parameters from our MCMC chains, we utilize the Python package \texttt{GetDist} \cite{2019arXiv191013970L}. Additionally, best-fit values for each cosmological model were obtained using Powell's 2009 BOBYQA bounded optimization algorithm, as implemented within \texttt{CosmoMC}. This procedure involved executing four independent minimizations from different random initial positions to ensure robust convergence.
 \section{Results and Observational Constraints}
\label{sec:results}
 
\begin{figure*}
\centering
\includegraphics[width=\textwidth]{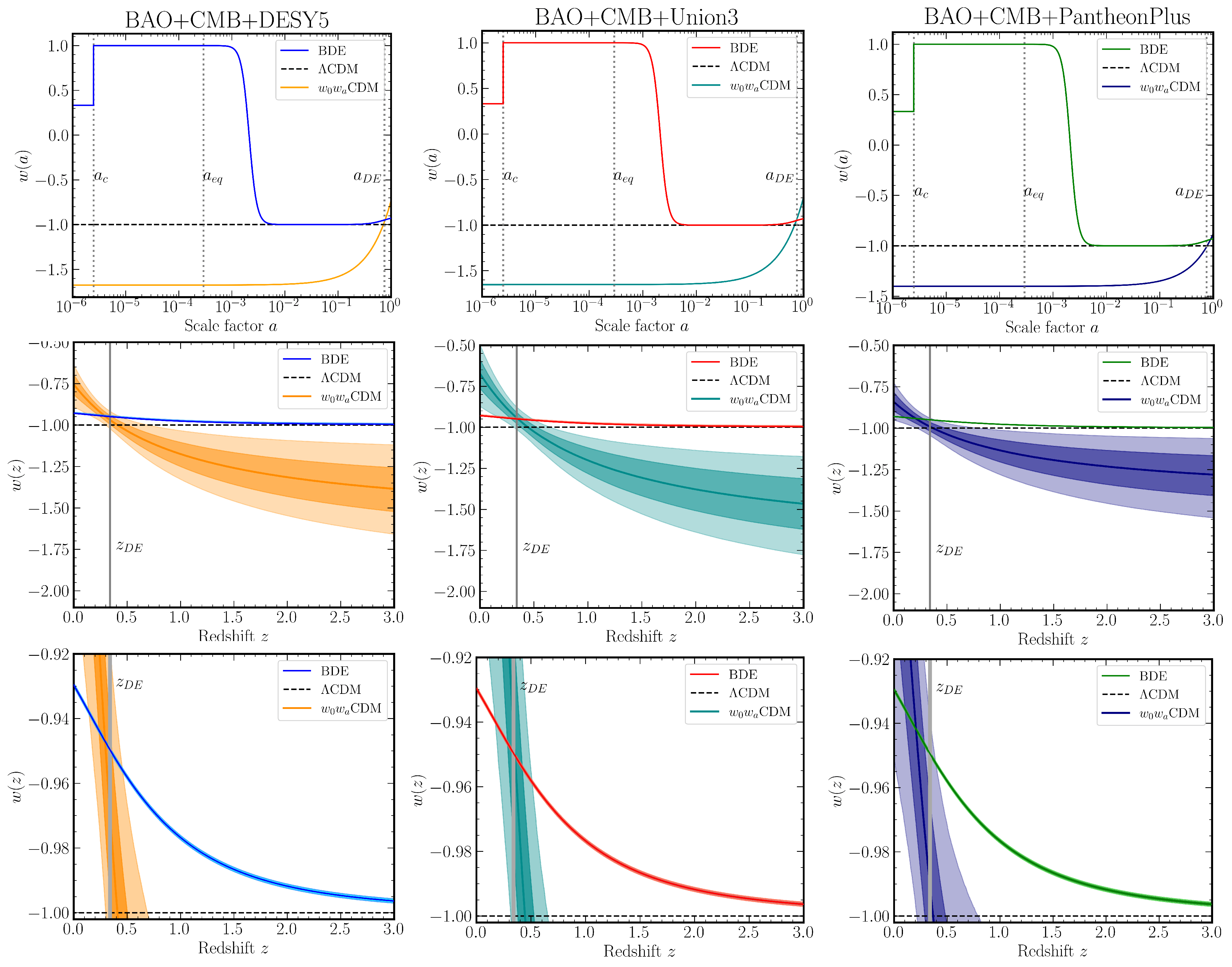}
\caption{\label{Fig_1:Equation_of_State_across_models} EoS $w(z)=P/\rho$ as function of redshift and scale factor in the framework of BDE-CDM, $\Lambda$CDM and $w_0w_a$CDM models from the joint analysis of BAO+CMB+DESY5 (left column), BAO+CMB+Union3 (middle column) and BAO+CMB+PantheonPlus (right column). In the first row, the evolution of $w(a)$ is exhibited as a function of scale factor $a$, from $a=1\times10^{-6}$ to present time $a_{0}=1$. In the second row, it shown the evolution of $w(z)$ in terms of redshift, in the redshift range $0<z<3$. In the third row, a closer look in the evolution of $w(z)$ in BDE-CDM model. The dotted and solid grey lines shows the scale factor $a_{c}$, $a_{eq}$, $a_{\mathrm{DE}}(z_{\mathrm{DE}})$ corresponding to the transition scale, radiation-matter equality and matter-dark energy equality.}
\end{figure*}
In this section, we present the observational constraints on three distinct DE models: the $\Lambda$CDM, the $w_0w_a$CDM, and BDE-CDM models. Our objective is to evaluate the statistical performance and cosmological constraints of each framework against the latest cosmological data. The models under consideration represent fundamentally different approaches to the dark energy problem. The $\Lambda$CDM and $w_0w_a$CDM models are primarily phenomenological. The latter, in particular, characterizes the evolution of the DE-EoS using the linear Chevallier-Polarski-Linder (CPL) parameterization \cite{PhysRevLett.90.091301, doi:10.1142/S0218271801000822}. This model has been central to recent investigations by DESI collaboration \cite{Adame_2025, tr6ykpc6}. Notably, recent DESI findings have suggested a potential deviation from a cosmological constant, excluding the $\Lambda$CDM model at a confidence level exceeding $95\%$,  points toward the necessity of dynamical dark energy. In contrast, the BDE model \cite{PhysRevD.72.043508, PhysRevLett.121.161303} provides a robust particle physics perspective, attributing the origin of DE to the condensation of a dark meson in the early universe. A defining characteristic of the BDE framework is its theoretical predictions on DE; unlike the $w_0w_a$CDM model, it introduces no free parameters within the DE sector. This lack of fine-tuning allows for specific predictions that can be tested directly against observational data.

\subsection{Evolution of the Equation of State} 
\begin{figure*}[hb]
\hspace{-1cm}
  \begin{tabular}{c@{\hspace{0.01em}}c@{\hspace{0.01em}}c}
      \includegraphics[width=5.5cm,height=4.5cm]{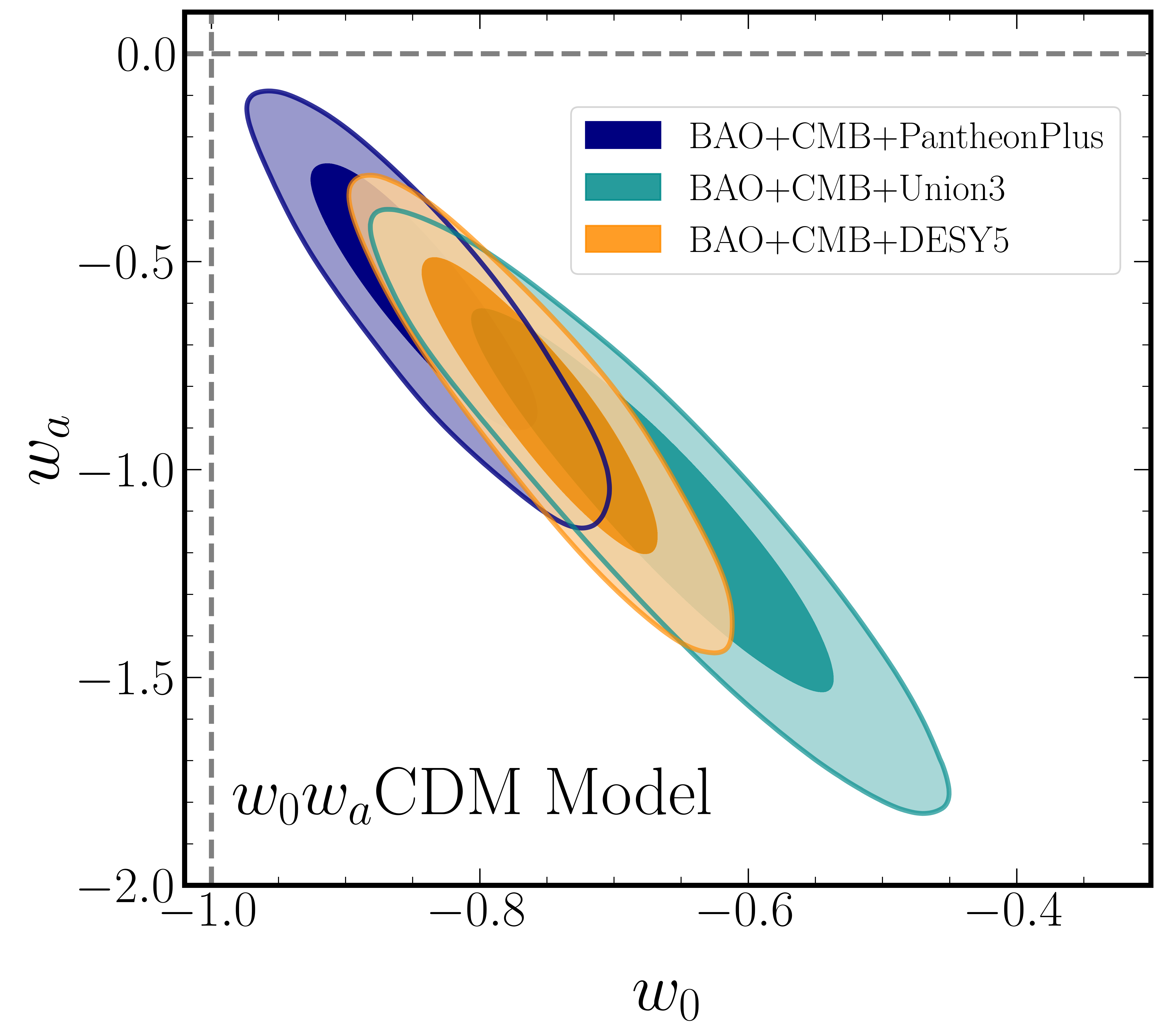} & 
      \includegraphics[width=5.5cm,height=4.5cm]{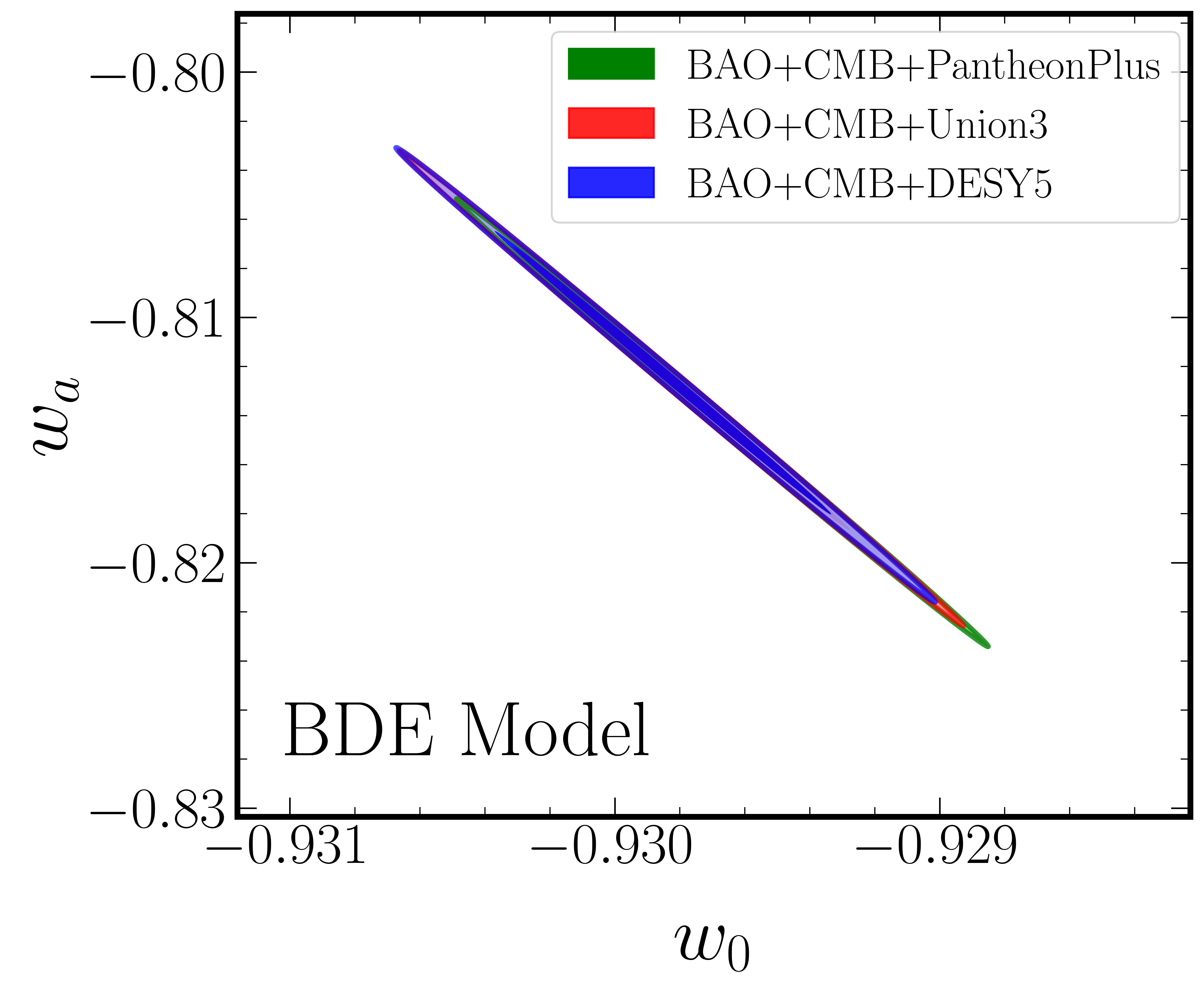}& 
      \includegraphics[width=5.5cm,height=4.5cm]{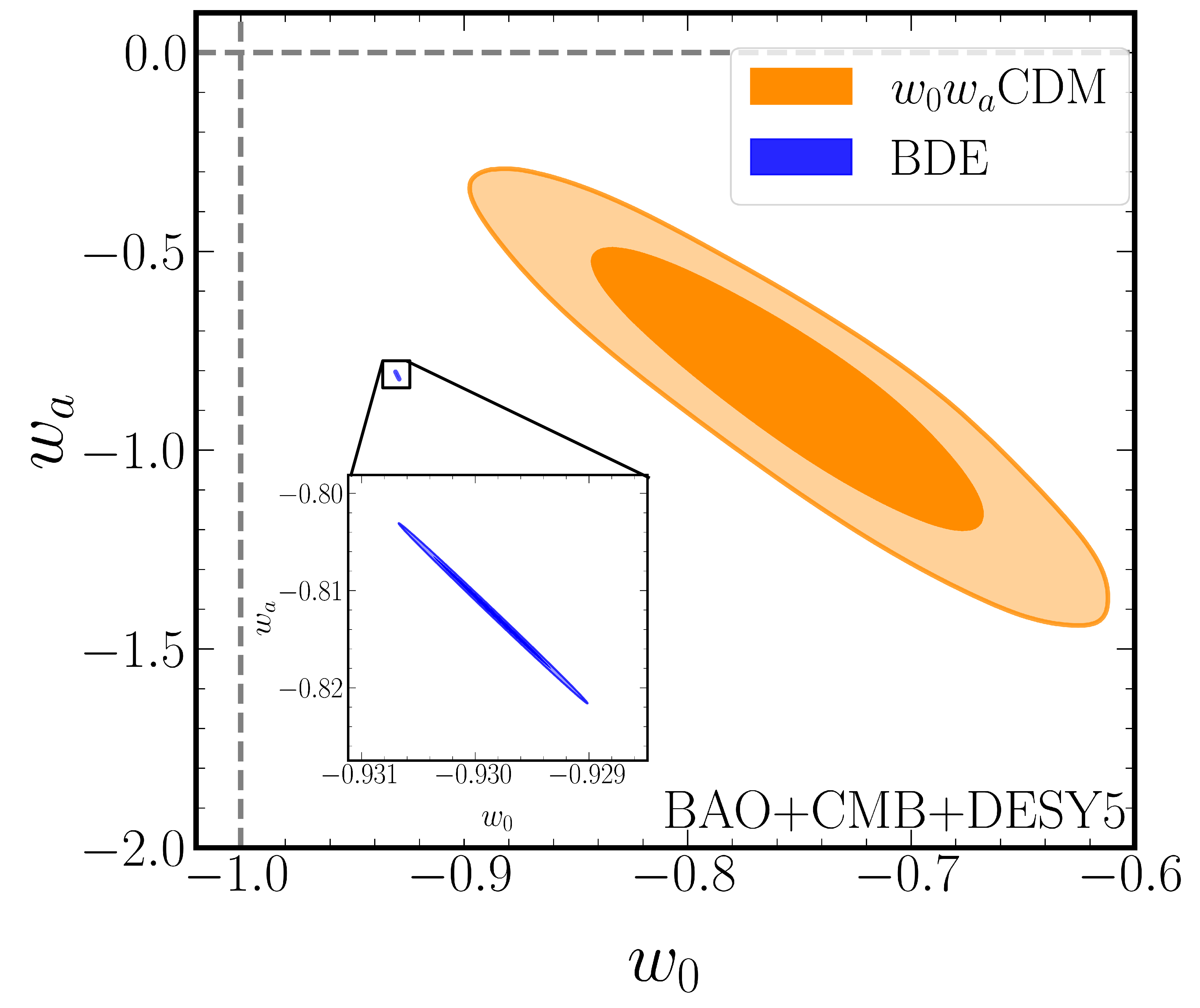}
  \end{tabular}
  \vspace*{8pt}
  \caption{\label{Fig_2: Cosmological constraints in w0wa plane} \textit{Left panel:} $68\%$ and $95\%$ marginalized posterior constraints in the $w_0$-$w_a$ plane for the flat $w_0w_a$CDM model, from DESI DR2 BAO combined with CMB and each of the PantheonPlus \cite{Scolnic_2022}, Union3 \cite{Rubin_2025}, and DESY5 \cite{Abbott_2024} SNe-Ia datasets. \textit{Middle panel:} $68\%$ add $95\%$ marginalized posterior constraints in the $w_0$-$w_a$ plane for the BDE-CDM model, exhibiting no shifting in the contour areas across the three data combinations considered. Each of these combinations sets a range of $w_0$ and $w_a$ of $-0.928<w_0<-0.931$, $-0.80<w_0<-0.83$, respectively. \textit{Right panel:} $68\%$ and $95\%$ marginalized posterior constraints comparative in the $w_0$-$w_a$ plane between BDE-CDM and $w_0w_a$CDM models, selecting the BAO+CMB+DESY5 data combination as a reference. }
\end{figure*}
It is presented in Figure \ref{Fig_1:Equation_of_State_across_models} a detailed comparison of the evolution of the DE-EoS, $w(z)=\rm P/\rho$, for the analyzed cosmological models: BDE-CDM, $w_0w_a$CDM, and $\Lambda$CDM. Each plot in Fig.\ref{Fig_1:Equation_of_State_across_models} exhibits error bands corresponding to the $68\%$ and $95\%$ confidence intervals, providing insight into the uncertainties associated with the $w_0,w_a$ parameters derived from joint analyses of BAO+CMB+PantheonPlus, BAO+CMB+Union3, and BAO+CMB+DESY5 data combinations. The limit imposed by the $\Lambda$CDM model is indicated by the dashed horizontal line. The first row of Fig.\ref{Fig_1:Equation_of_State_across_models} presents the evolution of the EoS for BDE, $\Lambda$CDM, and $w_0w_a$CDM models. The BDE model exhibits a consistent evolution in the EoS, $w(a)$, from early times to its present value, $w_0$, across all data combinations as seen in Fig.\ref{Fig_1:Equation_of_State_across_models}. 

The BDE-CDM model exhibits an interesting dynamical evolution of the EoS parameter $w_{\rm BDE}(a)$, which is characterized by distinct phases throughout cosmic history. At early times, prior to the condensation scale ($a < a_c$), the DGG consists of relativistic particles residing within the supersymmetric $SU(N_c=3)$ gauge group with $N_f = 6$ flavors. During this epoch, all DGG constituents are massless and relativistic, yielding $ w= 1/3, \text{with} \text{ } \rho_{\rm DGG}(a) = \rho_{\rm DGG}(a_c)\left(a_c/a\right)^4.$ The energy density dilutes as radiation, $\rho_{\rm DGG} \propto a^{-4}$, contributing an effective number of extra relativistic species $N_{\rm ext} = 0.945$~\cite{PhysRevD.99.103504}, which remains consistent with current observational bounds \cite{Mattias_Blennow_2012, MANGANO2011296, Saravanan_2025}. At the condensation phase transition ($a = a_c$), with $a_c = (2.489 \pm 0.007) \times 10^{-6}$ and $\Lambda_c = 43.93 \pm 0.13$ eV, based on the combination of DESI and CMB with DESY5 results as a reference, a non-perturbative phase transition occurs. The $SU(3)$ gauge coupling constant becomes strong, triggering the formation of neutral composite bound states—analogous to hadronization in QCD, where quarks confine into protons, neutrons, and mesons. At this transition, it is noticed the following aspects: The relativistic DGG constituents bind together, forming dark mesons; All energy stored in the DGG is completely transferred to the lightest meson field $\phi$, which constitutes our BDE particle: $\rho_{\rm DGG}(a_c) = \rho_{\rm BDE}(a_c) = 3\Lambda_c^4$; The extra relativistic species vanish: $N_{\rm ext} = 0$ for $a \geq a_c$ and the EoS undergoes a sharp transition from $w = 1/3$ to $w \simeq 1$. After condensation, in the range of $a_c < a \lesssim 1.1 \times 10^{-3}$, the scalar field $\phi$ evolves according to the Klein-Gordon equation in an expanding universe  $\ddot{\phi} + 3H\dot{\phi} + dV/d\phi = 0$, where the scalar potential is given by Eq.(\ref{eq1}). During this phase, the kinetic energy dominates over the potential energy, $\dot{\phi}^2/2 \gg V(\phi)$, resulting in $ w_{\rm BDE} = (\dot{\phi}^2/2 - V(\phi))/(\dot{\phi}^2/2 + V(\phi)) \simeq 1, \text{ } \rho_{\rm BDE} \propto a^{-6} $. This ``stiff'' phase causes the BDE energy density to dilute faster than any standard cosmological component—faster than radiation ($\propto a^{-4}$) or matter ($\propto a^{-3}$). This rapid dilution effectively ``hides'' the dark energy component during the radiation and early matter-dominated epochs, naturally explaining why dark energy becomes relevant only at late times without fine-tuning. Subsequently, in the range of $2 \times 10^{-3} \lesssim a \lesssim 2 \times 10^{-1}$, the universe continues to expand and Hubble friction—represented by the $3H\dot{\phi}$ term in the Klein-Gordon equation—progressively damps the field's kinetic energy. The field velocity decreases until the potential energy becomes dominant: $\dot{\phi}^2/2 \ll V(\phi) \text{ } \Rightarrow \text{ } \rho_{\rm BDE} \simeq V(\phi), \text{ } P_{\rm BDE} \simeq -V(\phi)$. Consequently, the equation of state approaches as $w_{\rm BDE} \simeq -1, \text{ } \rho_{\rm BDE} \simeq {\rm const.}$, resulting in a cosmological constant-like behavior. Finally, close to present time as visualized in the second and third rows of Fig.\ref{Fig_1:Equation_of_State_across_models}, a detailed comparison of the EoS for the BDE and $w_0w_a$CDM models is presented, highlighting their distinct dynamical behaviors. At the present epoch ($a_0 = 1$), we obtain $ w_0 = -0.9298 \pm 0.0003, \text{ } w_a \equiv -\left.dw/da\right|_{a=1} = -0.812 \pm 0.003$ for the combination of DESI and CMB with DESY5. The overall evolution of the EoS in BDE-CDM model satisfies $w > -1$ throughout cosmic history. In contrast, the $w_0w_a$CDM model demonstrates a transition into a phantom regime ($w(z) < -1$) in the distant past for $a < 0.8$, ultimately evolving to $w(z) > -1$ at the current time across all data sets considered which may lead to theoretical instabilities attributed to negative kinetic energy, which challenge the consistency of the underlying physical theories.

A clear distinction between the BDE-CDM and $w_0w_a$CDM models lies in the origin of the DE-EoS. In the $w_0w_a$CDM model employed by the DESI collaboration \cite{Adame_2025,tr6ykpc6}, the EoS is expressed as $w(a) = w_0 + w_a(1-a) $, where $w_0$ and $w_a$ are \textit{free parameters} constrained solely by observational data, without any underlying theoretical prediction for their values. On the other hand, the BDE-CDM model derives the values of $w_0$ and $w_a$, according to Eqs. (\ref{eq1}), (\ref{eq2}) and (\ref{eq:bde_acLc_theory}) as detailed in Section \ref{Theoretical Background}. Specifically, with $V(\phi)$, $\Lambda_c$, and $a_c$ fully specified, the Klein-Gordon equation coupled with the Friedmann equations, uniquely determines $\phi(a)$ and consequently $w(a)$ at all cosmic epochs. Therefore, $w_0$ and $w_a$ in the BDE-CDM framework are \textit{predicted values} of the model, obtained by evaluating the full $w(a)$ evolution at $a = 1$, $w_0$, and its first derivative $w_a$. They are not adjusted to fit data but emerge naturally from the underlying particle physics framework of our BDE-CDM model. This represents a qualitative departure from purely phenomenological approaches, where $w_0$ and $w_a$ serve as fitting parameters with no predictive theoretical basis. As demonstrated in the middle and right panels from Fig.~\ref{Fig_2: Cosmological constraints in w0wa plane}, the confidence contours in the $w_0$-$w_a$ plane for the BDE-CDM model are approximately $10{,}000$ times smaller in area compared to those of the $w_0w_a$CDM model. This reduction arises from the difference in parameter space dimensionality between the two models. The $w_0w_a$CDM model introduces three free parameters in the DE sector, namely $w_0$, $w_a$ and $\rho_{\rm DE}(a_0)$.  Observational data constrain this three-dimensional parameter volume, which when projected onto the $w_0$-$w_a$ plane yields extended confidence contours, they exhibit significant sensitivity to the choice of supernova dataset (PantheonPlus, Union3, or DESY5), reflecting the degeneracies inherent in parametrized models. In contrast, the BDE-CDM model contains zero free parameters in the DE sector since $\Lambda_c$ is predicted by gauge coupling unification (Eq.~\ref{eq2}), $a_c$ is derived from the constraint equation (Eq.~\ref{eq:bde_acLc_theory}) and $w_0$ and $w_a$ are derived from solving background evolution equations (see Section \ref{Theoretical Background} for details). As a result, we obtain small confidence contours in the $w_0$-$w_a$ plane for BDE-CDM model. 
 \begin{figure}[hb]
     \centering
     \includegraphics[width=0.8\textwidth]{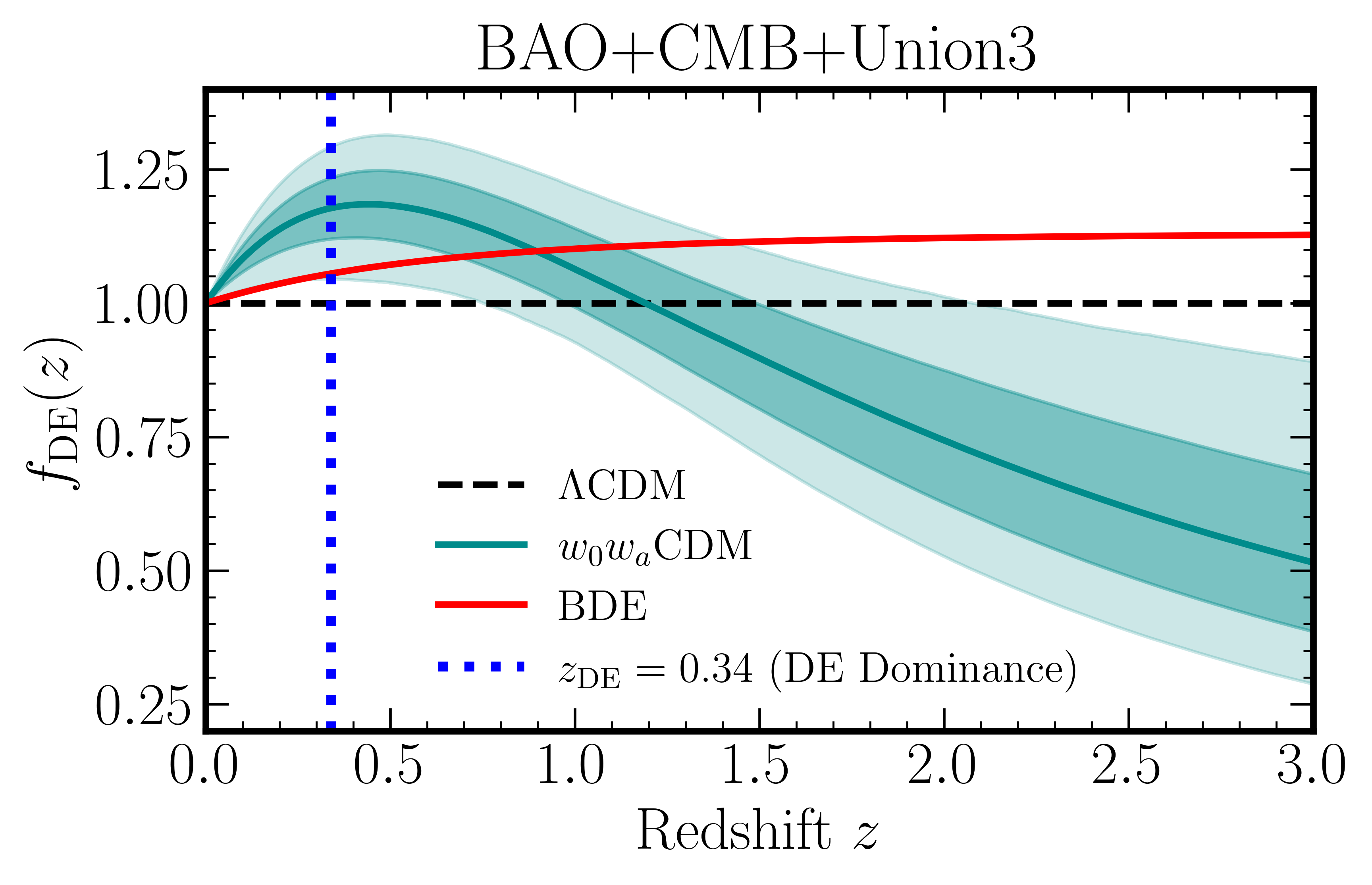}
     \caption{Normalized dark energy density, $f_{\mathrm{DE}}(z)\equiv \rho_{\mathrm{DE}}(z)/\rho_{\mathrm{DE}}(0)$, as a function of redshift in the framework of BDE-CDM and $w_0w_a$CDM model from the data combination BAO+CMB+Union3. The dotted vertical line indicate the dark energy-matter equality redshift. The horizontal dashed black line represents $\Lambda$CDM.}
     \label{Fig_3:normalized dark energy density}
 \end{figure}

A notable consequence of the BDE-CDM framework is the \textit{insensitivity} of the $w_0$-$w_a$ predictions to the choice of supernova compilation. As shown in Table~\ref{tab: Results of some parameters in the BDE, CPL, LCDM for the full datasets} and Fig.~\ref{Fig_2: Cosmological constraints in w0wa plane}, the BDE-CDM contours remain statistically identical across combinations with PantheonPlus, Union3, and DESY5 datasets. This stability contrasts sharply with the $w_0w_a$CDM results, where the central values shift substantially depending on the SNe-Ia sample employed. The robustness of BDE-CDM predictions stems directly from the theoretical determination of DE parameters, rendering them largely independent of late-time distance measurements. According to joint constraints from the combination of BAO, CMB and SNe-Ia probes, shown in the left and middle panel from Fig.~\ref{Fig_2: Cosmological constraints in w0wa plane}, we find that the results and the associated uncertainties do not vary depending on the choice of supernova dataset in the BDE-CDM model in comparison with $w_0w_a$CDM model. We find marginalized posterior means at a $68\%$ confidence level, the values for $w_0=w(a_0)$ and $w_a=-(\mathrm{d}w(a)/\mathrm{d}a)|_{a=a_{0}}$ predicted by BDE-CDM model are $w_0 = -0.9296 \pm 0.0003$, $w_a = -0.814 \pm 0.003$ from combination with PantheonPlus,  $w_0 = -0.9297 \pm 0.0003$, $w_a = -0.813 \pm 0.004$ from combination with Union3, and  $w_0 = -0.9299 \pm 0.0003$, $w_a = -0.812 \pm 0.003$ when using the DESY5 data. In contrast, for the $w_0w_a$CDM model, the values are $w_0 = -0.839 \pm 0.054$, $w_a = -0.61^{+0.23}_{-0.20}$ for the combination with PantheonPlus, $w_0 = -0.666 \pm 0.088$, $w_a = -1.08^{+0.32}_{-0.29}$ with Union3, and $w_0 = -0.752 \pm 0.058$, $w_a = -0.85^{+0.25}_{-0.22}$ with DESY5 consistent with \cite{tr6ykpc6}. In particular, these combinations of data favors a value of $w_0>-1$ and $w_a<0$, which reveals a significant discrepancy with cosmological constant EoS ($w_0=-1$, $w_a=0$) as reported by DESI collaboration \cite{Adame_2025, tr6ykpc6}. These findings indicate that the EoS predicted by the BDE-CDM model consistently satisfies $w>-1$ from early to present time. Conversely, the $w_0w_a$CDM model initially exhibits a phantom regime (where $w(z)<-1$) at high redshifts, subsequently transitioning to the $w(z)>-1$ domain at $z\lesssim 0.5$, as demonstrated in Fig.~\ref{Fig_1:Equation_of_State_across_models}. This evolution drives the $w_0w_a$CDM model into a phantom-like behavior. Such a transition yields to peculiar characteristics, as depicted in Fig. \ref{Fig_3:normalized dark energy density}, where the DE density, $\rho_{\mathrm{DE}}$, is normalized to its present-day value. In physical terms, a phantom EoS ($w(z)\leq-1$) corresponds to an energy density that increases with the expansion ($\mathrm{d}\rho_{\mathrm{DE}}/\mathrm{d}a>0$), reaching a maximum at $z\simeq0.45$—as inferred from the BAO+CMB+Union3 data in Fig.~\ref{Fig_3:normalized dark energy density}—when the EoS crosses the phantom divide ($w(z)=-1$), after which $\rho_{\mathrm{DE}}$ begins to decrease as the universe continues to expand. In contrast, within the BDE model, the DE density remains approximately constant ($\rho_{\mathrm{DE}}\simeq const.$) at high-redshift values and begins to decrease only at later times, as evidenced in Fig.~\ref{Fig_3:normalized dark energy density}.
\begin{figure*}[hb]
    \hspace{-1.2cm}
    \includegraphics[width=1.1\textwidth]{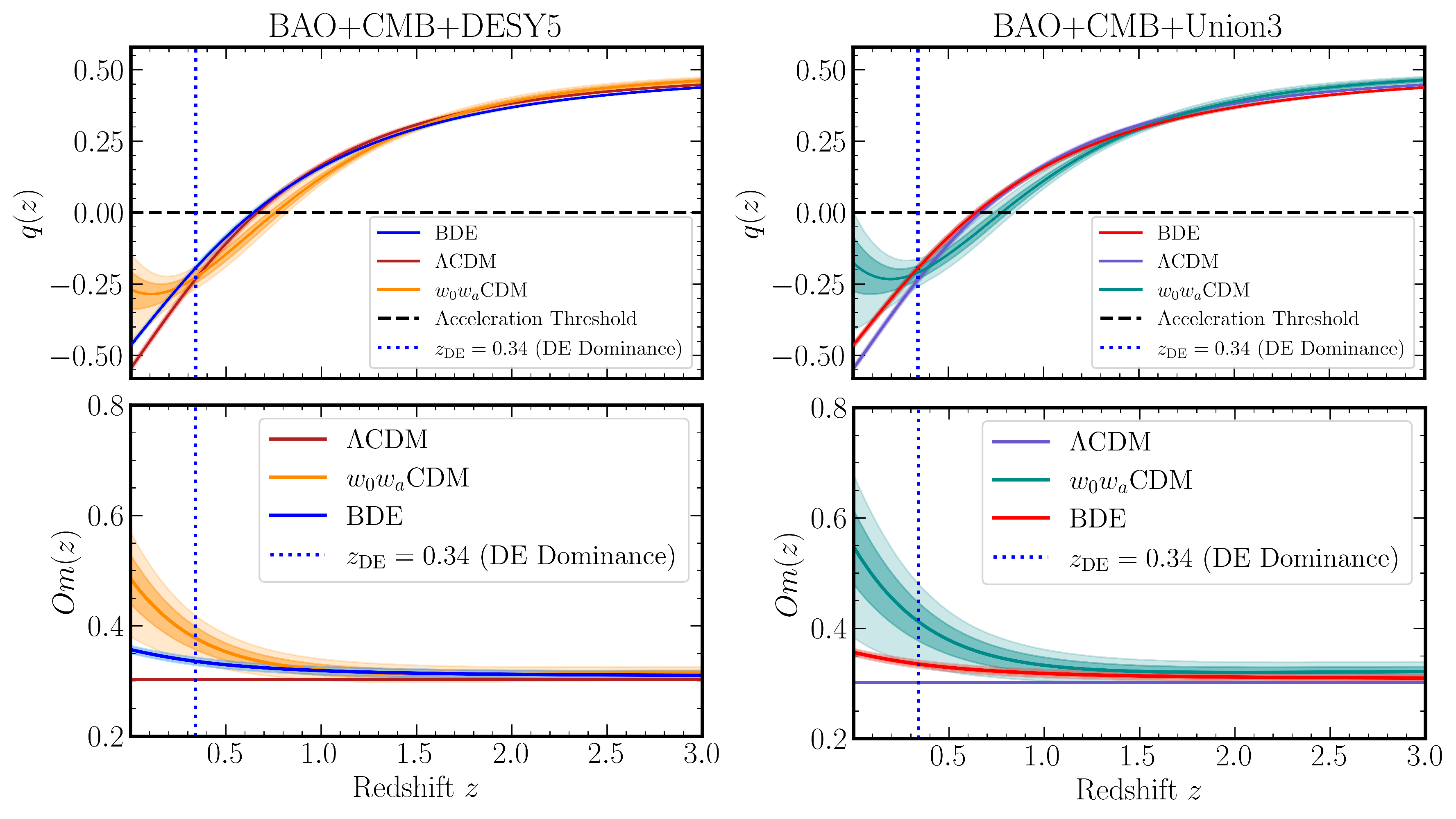}
    \caption{Evolution of the $\mathcal{O}m(z)$ diagnostic and deceleration parameter, $q(z)$, as a function of redshift for the BDE-CDM, $\Lambda$CDM and $w_0w_a$CDM models. The solid color lines of each model correspond to the median, $68$, and $95\%$ confidence levels obtained from the BAO+CMB+Union3 and BAO+CMB+DESY5 combinations. The blue dotted vertical lines indicate the dark energy-matter equality ($z_{\mathrm{DE}}$) redshift. }
    \label{fig_4: diagnostic and deceleration}
\end{figure*}

The cosmological models are evaluated using two key quantity terms, which serve as sensitive probes of new physics because they depend solely on the shape of the normalized expansion history $H(z)/H_{0}$. One such diagnostic, $Om(z)$, is sensitive to the DE-EoS and functions as a null test for the $\Lambda$CDM model. Specifically, $Om(z)$ is a geometric diagnostic that relates the Hubble parameter, which quantifies the universe's expansion rate, to redshift. This property enables $Om(z)$ to distinguish between dynamical dark energy (DDE) models and $\Lambda$CDM, independent of the matter density parameter $\Omega_{m}$. A constant $Om(z)$ across redshifts indicates that DE behaves as a cosmological constant $\Lambda$. In contrast, a positive slope in $Om(z)$ suggests phantom dark energy ($w < -1$), while a negative slope indicates quintessence ($w > -1$). Consistent with \cite{PhysRevD.78.103502, PhysRevLett.101.181301}, $Om(z)$ for a spatially flat universe is defined as follows:
\begin{equation}
    Om(z) \equiv \frac{H^{2}(z)/H^{2}_{0}-1}{(1+z)^{3}-1}.
\end{equation}
Here, $H_0$ denotes the present value of the Hubble parameter. The reconstructed $Om(z)$, as shown in the lower plots of Fig.~\ref{fig_4: diagnostic and deceleration}, demonstrates a significant $(>2\sigma)$ deviation of the $w_0w_a$CDM and BDE-CDM models from the cosmological constant $\Lambda$ model within the ranges $0\lesssim z \lesssim 0.5$ and $0 \lesssim z \lesssim 1.5$, respectively, for both BAO+CMB+DESY5 and BAO+CMB+Union3 data combinations. The firebrick and slate-blue lines indicate the best-fit $\Lambda$CDM value of $\Omega_m=0.308$, illustrating that $Om(z)=\Omega_{m0}$ for $\Lambda$CDM, while $Om(z)>\Omega_{m0}$ for quintessence $(w>-1)$ and $Om(z)<\Omega_{m0}$ for phantom $(w<-1)$ models. This observed effect may suggest that dark energy is decaying into an unknown component. A similar trend in $Om(z)$ was previously reported with DESI DR1 and is also evident in the current DR2 data. On the other hand, the deceleration parameter $q(z)$ is defined as 
\begin{equation}
    q(z)\equiv -\frac{\ddot{a}a}{\dot{a}^{2}}=\frac{\mathrm{d}\text{ }\mathrm{ln}H}{\mathrm{d}\text{ }\mathrm{ln}(1+z)}-1,
\end{equation}
where it tracks the logarithmic derivative of $H(z)/H_{0}$. The deceleration parameter, $q(z)$, reveals that cosmic acceleration ($q<0$) within the $w_0w_a$CDM framework started at an earlier epoch than previously anticipated as seen in the upper panels from Fig.~\ref{fig_4: diagnostic and deceleration} for both BAO+CMB+DESY5 and BAO+CMB+Union3 data combinations. Specifically, the onset of acceleration occurs at a redshift of $z\simeq0.8$ according to $w_0w_a$CDM model, which precedes the transition redshift predicted by both BDE-CDM and $\Lambda$CDM models (at $z\simeq 0.65$), with evidence for a subsequent deceleration at later times. Notably, this behavior in $q(z)$ is corroborated by recent analyses conducted by the DESI collaboration using DESI DR2 data \cite{tr6ykpc6}.

\begin{table}[htpb]
\hspace*{-1cm}
\begin{tabular}{lcccccc}
\hline
& \multicolumn{2}{c}{BDE-CDM} & \multicolumn{2}{c}{$w_0w_a$CDM} & \multicolumn{2}{c}{$\Lambda$CDM}\\
\hline
Parameters & Best fit & $68\%$ limits & Best fit & $68\%$ limits & Best fit & $68\%$ limits\\ \hline
& \multicolumn{1}{c}{ } & \multicolumn{3}{c}{\textbf{BAO+CMB+PantheonPlus}} & \multicolumn{2}{c}{ }\\
\hline
$10^{6} a_{c}$      & 2.481   & 2.486 $\pm$ 0.007  &    ---     &   --- & ---   &  ---  \\
 $\Lambda_{c}$[eV ] & 44.08   & 44.00 $\pm$ 0.13   &    ---     &   --- &  ---  & ---    \\
 $\Omega_{b}h^2$    & 0.02273 & 0.02265 $\pm$ 0.00012 & 0.02243 &  0.02242 $\pm$ 0.00013 & 0.02255 & 0.02253 $\pm$ 0.00012 \\
 $\Omega_{c}h^2$    & 0.1174  & 0.1177 $\pm$ 0.0006   & 0.1191  & 0.1190   $\pm$ 0.0009  & 0.1174 & 0.1175 $\pm$ 0.0006   \\
 $\mathrm{ln}(10^{10}A_{s})$ & 3.132 & 3.050 $\pm$ 0.016 & 3.047 &  3.045 $\pm$ 0.016 & 3.043 & 3.047 $\pm$ 0.016     \\
$\tau$              & 0.098    & 0.058 $\pm$ 0.008 & 0.050   & 0.056 $\pm$ 0.007 &  0.057   & 0.058 $\pm$ 0.008    \\
 $n_{s}$            & 0.9783   & 0.9753 $\pm$ 0.0034 & 0.9676 & 0.9671 $\pm$ 0.0037 & 0.9716 & 0.9707  $\pm$ 0.0033     \\
 $100\theta_{MC}$   & 1.04115  & 1.04110 $\pm$ 0.00027 & 1.04103  & 1.04105 $\pm$ 0.00028 & 1.04123 & 1.04123 $\pm$ 0.00027 \\
 $w_{0}$            & -0.9294  & -0.9296 $\pm$ 0.0003  & -0.8810 & -0.8390 $\pm$ 0.0542  &  -1      &  -1    \\
 $w_{a}$            & -0.816  & -0.814 $\pm$ 0.003  & -0.580 & -0.610$^{+0.23}_{-0.20}$   &  0       & 0    \\
 $H_{0}$            & 67.81    & 67.64  $\pm$ 0.27     & 67.23   & 67.50 $\pm$ 0.64  & 68.50    & 68.46 $\pm$ 0.30 \\
 $\Omega_{m}$       & 0.3047   & 0.3067 $\pm$ 0.0036    & 0.3055 & 0.3038 $\pm$ 0.0068 & 0.2997 & 0.3002 $\pm$ 0.0038 \\
 $\Omega_{\mathrm{DE}}$ & 0.6952 & 0.6932 $\pm$ 0.0036  & 0.6944 & 0.6961 $\pm$ 0.0068 & 0.7002 & 0.6997 $\pm$ 0.0038 \\
 $\sigma_{8}$       & 0.8412    & 0.8080 $\pm$ 0.0071 & 0.8140   & 0.8172 $\pm$ 0.0118 & 0.8037 & 0.8051 $\pm$ 0.0072 \\
$r_{\mathrm{drag}}\mathrm{[Mpc]}$ & 147.26 & 147.27 $\pm$ 0.19  & 147.23  & 147.28 $\pm$ 0.24 & 147.55 & 147.56 $\pm$ 0.19 \\
\hline
& \multicolumn{1}{c}{ } & \multicolumn{3}{c}{\textbf{BAO+CMB+Union3}} & \multicolumn{2}{c}{ }\\
 \hline
$10^{6} a_{c}$      & 2.483   & 2.488 $\pm$ 0.008 &    ---     &          ---          & ---   &  ---  \\
 $\Lambda_{c}$[eV ] & 44.04 & 43.95 $\pm$ 0.14  &    ---     &          ---          &  ---  & ---    \\
 $\Omega_{b}h^2$    & 0.02272 & 0.02263 $\pm$ 0.00012 & 0.02238 &  0.02239 $\pm$ 0.00013 & 0.02252 & 0.02251 $\pm$ 0.00012 \\
 $\Omega_{c}h^2$    & 0.1176 & 0.1179 $\pm$ 0.0006 & 0.1195 &  0.1195  $\pm$ 0.0009  & 0.1177 & 0.1177 $\pm$ 0.0006 \\
 $\mathrm{ln}(10^{10}A_{s})$ & 3.135 & 3.050 $\pm$ 0.016 & 3.033 & 3.044 $\pm$ 0.015  & 3.045 & 3.046 $\pm$ 0.016  \\
$\tau$              & 0.100 & 0.058 $\pm$ 0.007 & 0.047 & 0.055 $\pm$ 0.007  & 0.056 & 0.057 $\pm$ 0.008    \\
 $n_{s}$            & 0.9781 & 0.9746 $\pm$ 0.0033 & 0.9667 & 0.9659 $\pm$ 0.0037 & 0.9709 & 0.9702 $\pm$ 0.0033     \\
 $100\theta_{MC}$   & 1.04113 & 1.04109 $\pm$ 0.00027  & 1.04096 &  1.04099 $\pm$ 0.00028 & 1.04120 & 1.04120 $\pm$ 0.00027 \\
 $w_{0}$            & -0.9295 & -0.9297 $\pm$ 0.0003 & -0.7056 &  -0.6667 $\pm$ 0.0887  &  -1      &  -1    \\
 $w_{a}$            & -0.815 & -0.812 $\pm$ 0.003 & -0.974  & -1.084$^{+0.32}_{-0.29}$ &  0     & 0    \\
 $H_{0}$            & 67.75 & 67.56 $\pm$ 0.28 & 66.34 & 66.03 $\pm$ 0.83 & 68.38 & 68.35 $\pm$ 0.29 \\
 $\Omega_{m}$       & 0.3057 & 0.3079 $\pm$ 0.0038 & 0.3239 & 0.3270 $\pm$ 0.0086 & 0.3012 & 0.3016 $\pm$ 0.0037 \\
 $\Omega_{\mathrm{DE}}$ & 0.6942 & 0.6920 $\pm$ 0.0038 & 0.6760 & 0.6729 $\pm$ 0.0086 & 0.6987 & 0.6983 $\pm$ 0.0037 \\
 $\sigma_{8}$       & 0.8428 & 0.8086 $\pm$ 0.0072 & 0.7995 & 0.8009 $\pm$ 0.0116 & 0.8051 & 0.8055 $\pm$ 0.0071 \\
$r_{\mathrm{drag}}\mathrm{[Mpc]}$ & 147.23  & 147.24 $\pm$ 0.19 & 147.18 & 147.18 $\pm$ 0.23 & 147.51 & 147.52 $\pm$ 0.19 \\
\hline
& \multicolumn{1}{c}{ } & \multicolumn{3}{c}{\textbf{BAO+CMB+DESY5}} & \multicolumn{2}{c}{ }\\
 \hline
 $10^{6} a_{c}$     & 2.483 & 2.489 $\pm$ 0.007 &    ---     &          ---          & ---   &  ---  \\
 $\Lambda_{c}$[eV ] & 44.04 & 43.93 $\pm$ 0.13 &    ---     &          ---          &  ---  & ---    \\
 $\Omega_{b}h^2$    & 0.02273 & 0.02262 $\pm$ 0.00012 & 0.02239 & 0.02240 $\pm$ 0.00013 & 0.02250 & 0.02249 $\pm$ 0.00012    \\
 $\Omega_{c}h^2$    & 0.1176 & 0.1180 $\pm$ 0.0006 & 0.1198 &  0.1193  $\pm$ 0.0009  & 0.1179 & 0.1180 $\pm$ 0.0006   \\
 $\mathrm{ln}(10^{10}A_{s})$& 3.160 & 3.050 $\pm$ 0.016 & 3.038 & 3.044 $\pm$ 0.016 & 3.044 & 3.046 $\pm$ 0.016  \\
$\tau$              & 0.112 & 0.058 $\pm$ 0.008 & 0.050 & 0.055 $\pm$ 0.007 & 0.058 & 0.057 $\pm$ 0.007  \\
 $n_{s}$            & 0.9786 & 0.9744 $\pm$ 0.0033 & 0.9662 & 0.9662 $\pm$ 0.0037 & 0.9712 & 0.9695 $\pm$ 0.0034  \\
 $100\theta_{MC}$   & 1.04113 & 1.04107 $\pm$ 0.00027 & 1.04098 & 1.04100 $\pm$ 0.00028 & 1.04119 & 1.04118 $\pm$ 0.00027 \\
 $w_{0}$            & -0.9295 & -0.9298 $\pm$ 0.0003 & -0.7363 & -0.7521 $\pm$ 0.0581  &  -1      &  -1    \\
 $w_{a}$            & -0.8150 & -0.812 $\pm$ 0.003 & -0.939 & -0.857$^{+0.25}_{-0.22}$ &  0     & 0    \\
 $H_{0}$            & 67.74 & 67.51 $\pm$ 0.27 & 66.84 & 66.85 $\pm$ 0.56  & 68.27 & 68.24 $\pm$ 0.29 \\
 $\Omega_{m}$       & 0.3058 & 0.3086 $\pm$ 0.0037 & 0.3196  & 0.3186 $\pm$ 0.0056 & 0.3026 & 0.3030 $\pm$ 0.0037 \\
 $\Omega_{\mathrm{DE}}$ & 0.6941 & 0.6913 $\pm$ 0.0037 & 0.6803 & 0.6813 $\pm$ 0.0056  & 0.6973 & 0.6969 $\pm$ 0.0037 \\
 $\sigma_{8}$       & 0.8536 & 0.8090 $\pm$ 0.0071 & 0.8075 & 0.8067 $\pm$ 0.0110  & 0.8057 & 0.8064 $\pm$ 0.0071 \\
$r_{\mathrm{drag}}\mathrm{[Mpc]}$ & 147.20  & 147.22 $\pm$ 0.20 & 147.11 & 147.22 $\pm$ 0.23 & 147.48 & 147.47 $\pm$  0.20\\\hline
\end{tabular}
\caption{\label{tab: Results of some parameters in the BDE, CPL, LCDM for the full datasets} Best-fit, 68\%, 95\% cosmological results from the joint analysis of the DESI DR2 BAO \cite{tr6ykpc6}, CMB  \cite{refId0} with PantheonPlus \cite{Scolnic_2022}, Union3 \cite{Rubin_2025} and DESY5 \cite{Abbott_2024} for BDE-CDM, $\Lambda$CDM and $w_0w_a$CDM models.}
\end{table}
\subsection{Cosmological Constraints}
 \begin{figure*}[ht!]
 \centering
\includegraphics[width=\textwidth]{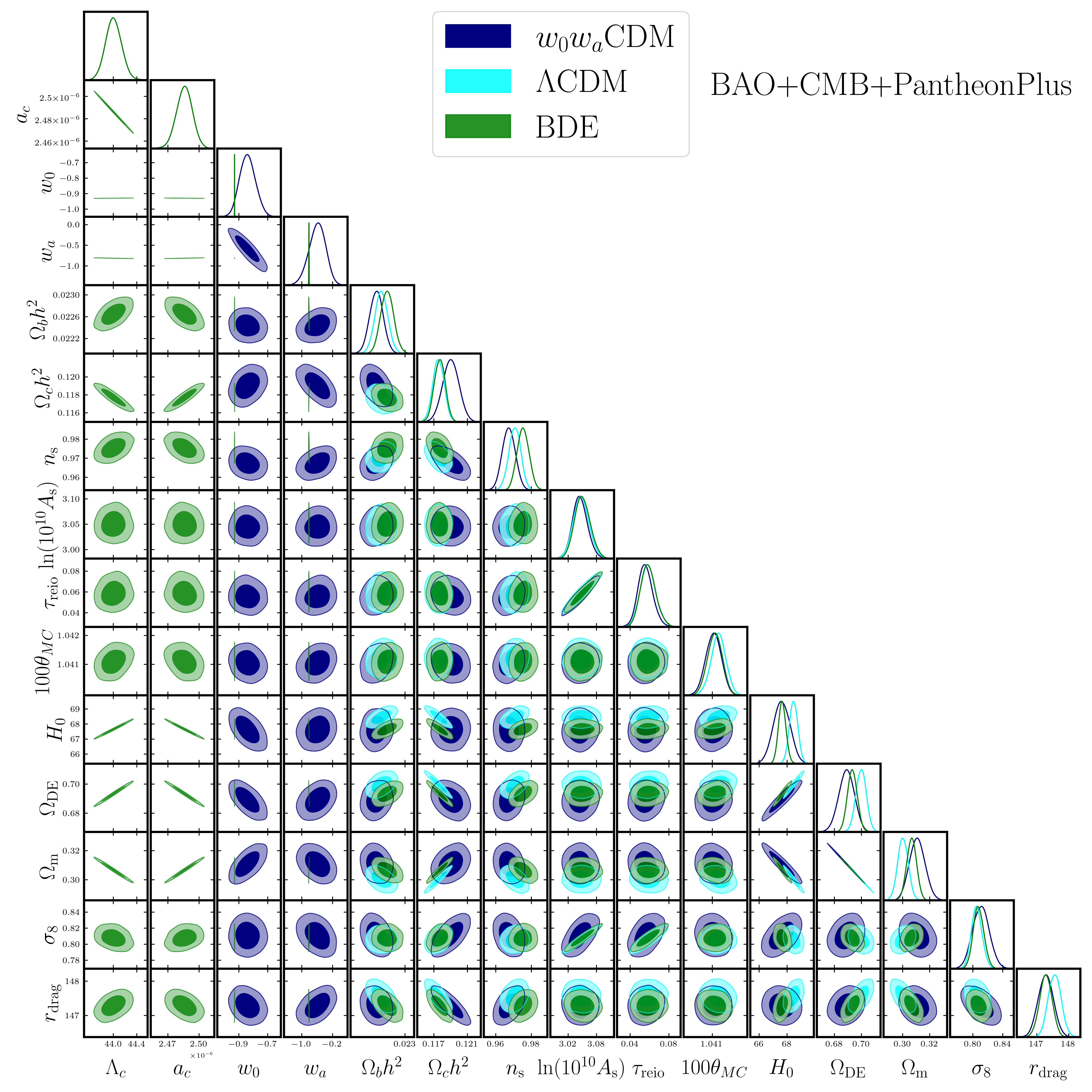}
\caption{\label{Fig_5: cosmological constraints BAO+CMB+PantheonPlus} Marginalized distributions and $68\%$ and $95\%$ confidence contours of some cosmological key parameters for the BDE-CDM (green), $\Lambda$CDM (aqua blue), and $w_0w_a$CDM (navy blue) models from the combination of DESI DR2 BAO \cite{tr6ykpc6}, CMB \cite{refId0}, and PantheonPlus \cite{Scolnic_2022}. }
\end{figure*}
 \begin{figure*}[ht!]
 \centering
\includegraphics[width=\textwidth]{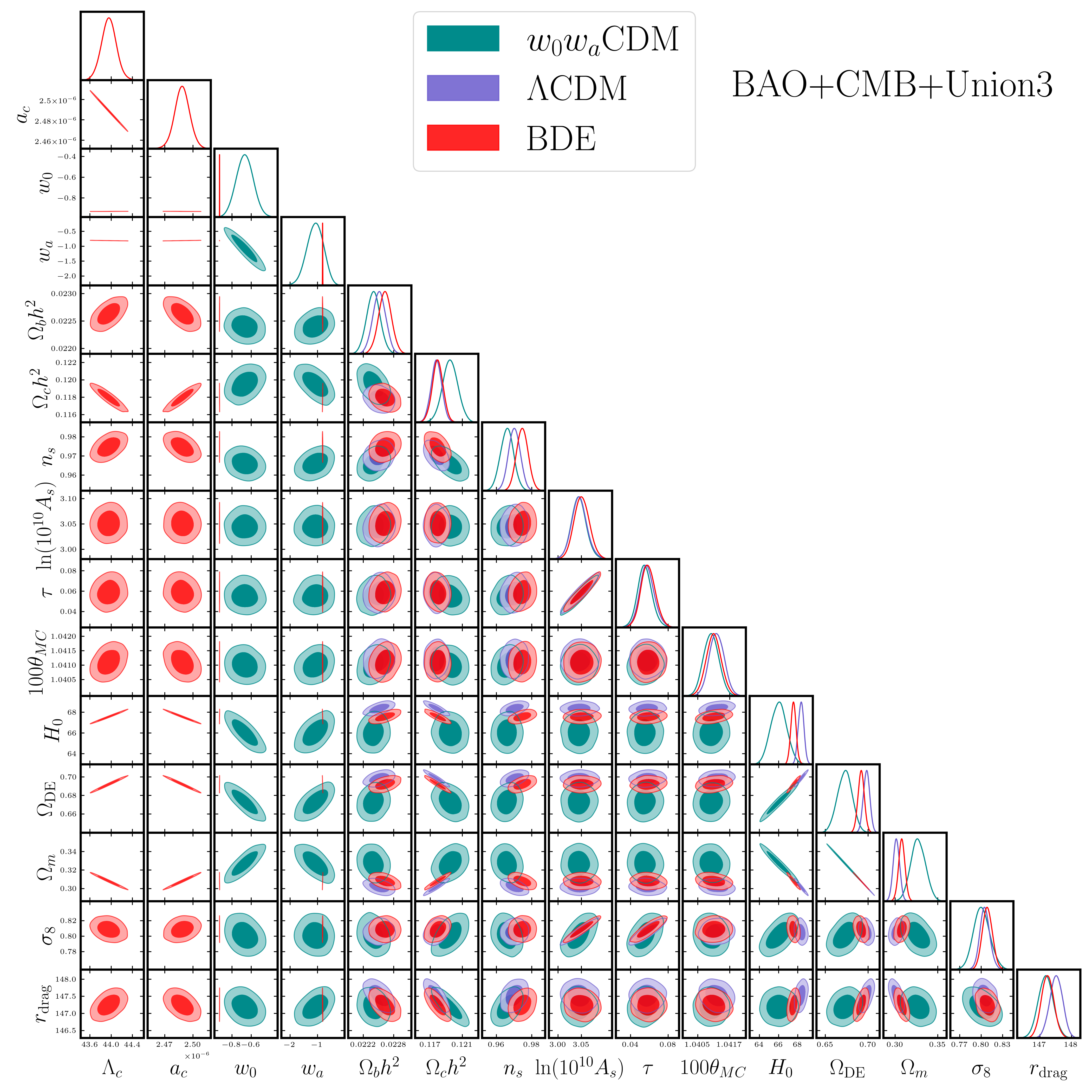}
\caption{\label{Fig_6: cosmological constraints BAO+CMB+Union3} Marginalized distributions and $68\%$ and $95\%$ confidence contours of some cosmological key parameters for the BDE-CDM (red), $\Lambda$CDM (slate blue), and $w_0w_a$CDM (dark cyan) models from the combination of DESI DR2 BAO \cite{tr6ykpc6}, CMB \cite{refId0}, and Union3 \cite{Rubin_2025}. }
\end{figure*}
 \begin{figure*}[ht!]
 \centering
\includegraphics[width=\textwidth]{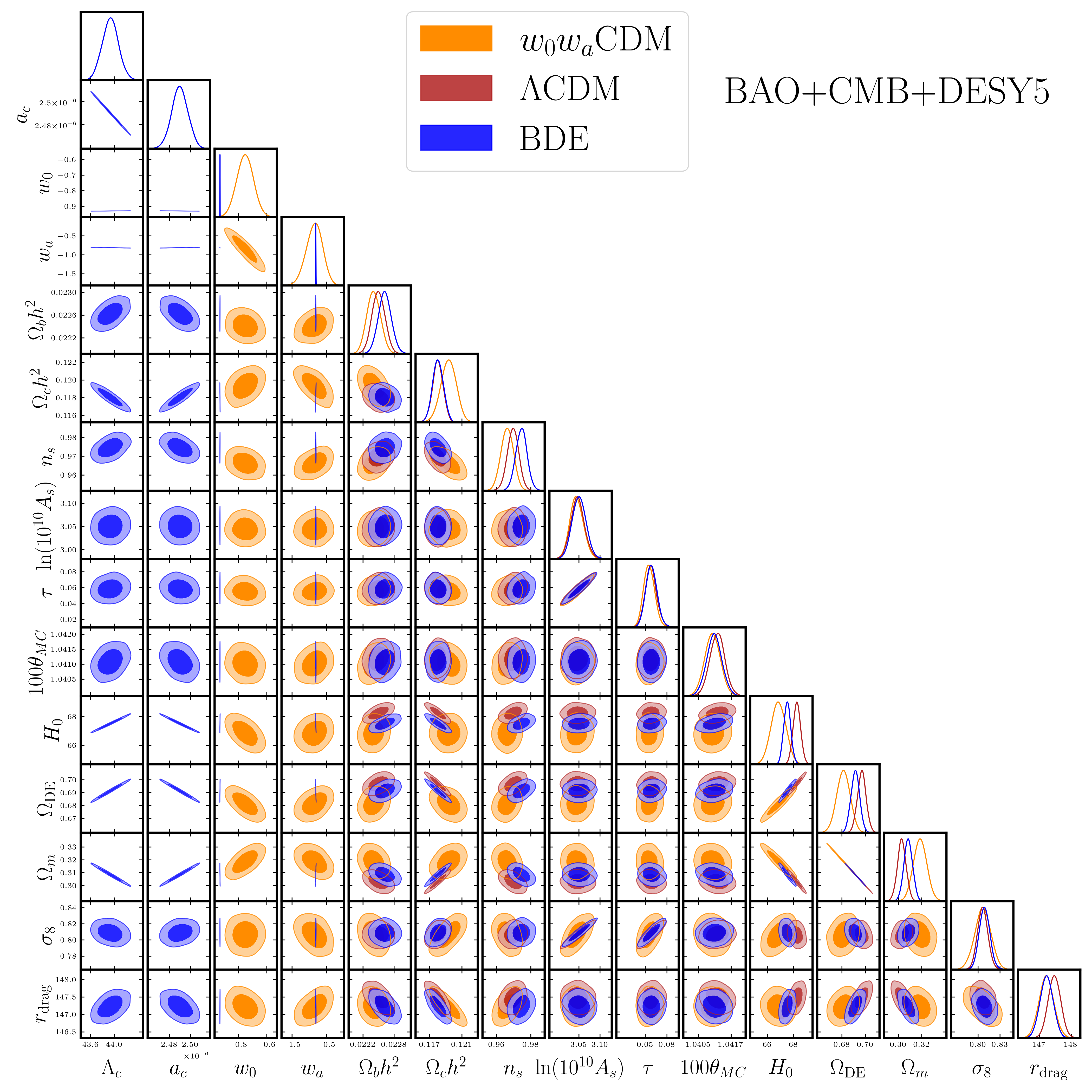}
\caption{\label{Fig_7: cosmological constraints BAO+CMB+DESY5}  Marginalized distributions and $68\%$ and $95\%$ confidence contours of some cosmological key parameters for the BDE (blue), $\Lambda$CDM (firebrick), and $w_0w_a$CDM (dark orange) models from the combination of DESI DR2 BAO \cite{tr6ykpc6}, CMB \cite{refId0}, and DESY5 \cite{Abbott_2024}. }
\end{figure*}

 Marginalized constraints on key cosmological parameters at the 68\% and 95\% confidence levels are rigorously quantified for the BDE-CDM model, as well as for the $w_0w_a$CDM and $\Lambda$CDM models. These constraints are derived from a comprehensive analysis that combines the DESI DR2 BAO dataset \cite{tr6ykpc6} with CMB measurements \cite{refId0} and supernovae Type Ia (SNe-Ia) samples from PantheonPlus \cite{Scolnic_2022}, Union3 \cite{Rubin_2025}, and DESY5 \cite{Abbott_2024}. The resulting marginalized posterior distributions and their confidence contours are visually depicted in Figs. \ref{Fig_5: cosmological constraints BAO+CMB+PantheonPlus}, \ref{Fig_6: cosmological constraints BAO+CMB+Union3}, and \ref{Fig_7: cosmological constraints BAO+CMB+DESY5}. The central (mean) parameter estimates at the 68\% confidence level and the corresponding best-fit values for each cosmological model are detailed in Table \ref{tab: Results of some parameters in the BDE, CPL, LCDM for the full datasets}. Notably, the constraints on the six fundamental $\Lambda$CDM parameters—namely, the baryon density $\Omega_{b}h^{2}$, cold dark matter density $\Omega_{c}h^{2}$, acoustic scale $100\theta_{MC}$, reionization optical depth $\tau$, amplitude of scalar fluctuations $\mathrm{ln}(10^{10}A_{s})$, and scalar spectral index $n_{s}$—as well as the derived parameters ($H_0$, $\Omega_{m}$, $\Omega_{\mathrm{DE}}$, $r_{\mathrm{drag}}$, and $\sigma_{8}$), exhibit strong concordance across the $\Lambda$CDM, $w_0w_a$CDM, and BDE-CDM models at the 1$\sigma$ confidence level. This consistency is substantiated by the results presented in Table \ref{tab: Results of some parameters in the BDE, CPL, LCDM for the full datasets} and the aforementioned figures. Constraints in the $\Omega_{m}$-$H_{0}$ plane for the BDE-CDM and $\Lambda$CDM models, derived from the combination of DESI BAO DR2 measurements with CMB and each SNe-Ia sample, indicate higher values of $H_0$ and lower values of $\Omega_{m}$. Specifically, $H_{0}=67.56\pm0.28$ $\mathrm{km/s/Mpc}$ and $\Omega_{m}=0.3079\pm0.0038$ are found for BDE-CDM model, while $H_{0}=68.35\pm0.29$ $\mathrm{km/s/Mpc}$ and $\Omega_{m}=0.3016\pm0.0037$ are obtained for $\Lambda$CDM, based on DESI DR2 BAO combined with CMB and Union3. In contrast, the $w_0w_a$CDM model yields $H_0=66.03\pm0.83$ $\mathrm{km/s/Mpc}$ and $\Omega_{m}=0.3270\pm0.0086$. Comparable results are observed with the other SNe-Ia datasets, as detailed in Table \ref{tab: Results of some parameters in the BDE, CPL, LCDM for the full datasets}. The values of $H_{0}$ within the BDE-CDM model across the three combinations of datasets are clearly more in agreement with the value obtained by \textit{Planck} \cite{refId0} than with those from the Cepheid-calibrated local distance ladder \cite{Riess_2022, Riess_2019}. 
 As for the BDE-CDM key parameters, the determined value for transition scale factor and condensation energy scale, $a_{c}=(2.486 \pm 0.007)\times10^{-6}$, $\Lambda_{c} = 44.00 \pm 0.13$ eV for BAO+CMB+PantheonPlus, $a_{c}=(2.488 \pm 0.008)\times10^{-6}$, $\Lambda_{c} = 43.95 \pm 0.14$ eV for BAO+CMB+Union3 and $a_{c}=(2.489 \pm 0.007)\times10^{-6}$, $\Lambda_{c} = 44.93 \pm 0.13$ eV for BAO+CMB+DESY5, are found to agree with the theoretically predicted range $\Lambda_c^{\text{th}} = 34^{+16}_{-11}$ eV (68\% credible interval), which is determined by BDE constraints. Overall, the integration of DESI DR2 BAO, CMB, and SNe-Ia datasets makes robust constraints on fundamental cosmological parameters, thereby advancing our understanding of the universe’s composition, expansion history, and the nature of its initial conditions.
\begin{figure*}
  \centering
  \begin{tabular}{c@{\hspace{1em}}c}
      \includegraphics[width=\textwidth]{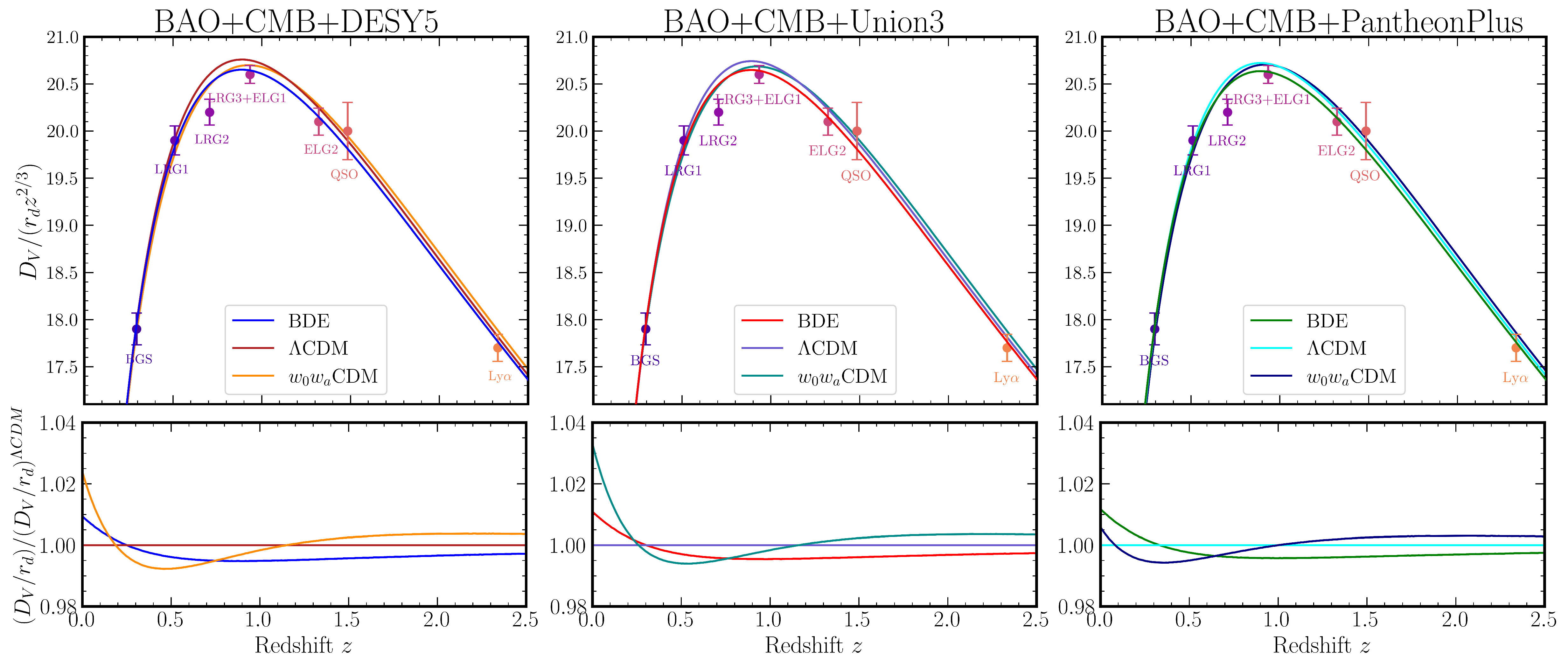}\\ 
     \includegraphics[width=\textwidth]{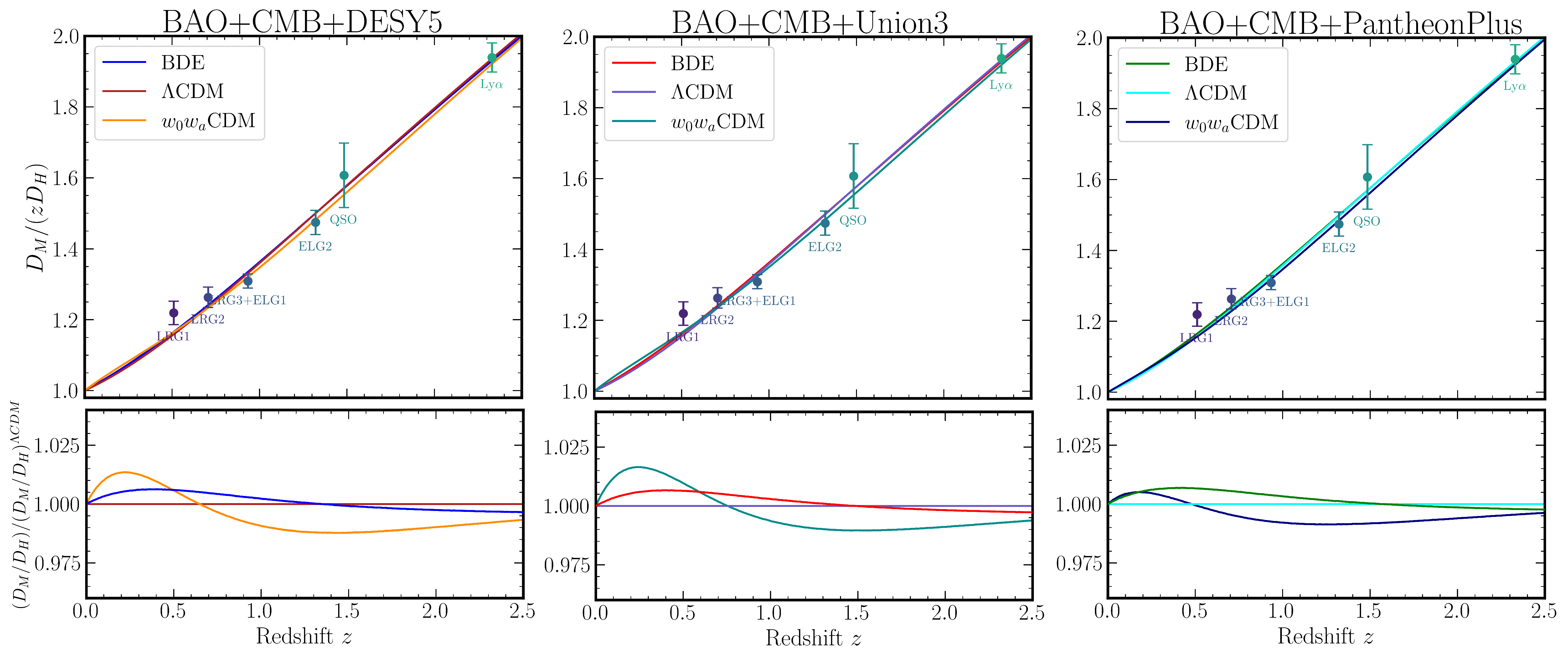}
  \end{tabular}
  \vspace*{8pt}
  \caption{\label{Fig_8: Distance redshift measurements} The DESI project measures the Baryon Acoustic Oscillation (BAO) distance scales at different redshifts. These measurements compare the angle-averaged distance, defined as $D_{V} \equiv (zD^{2}_M D_H)^{1/3}$ (for clarity, an arbitrary scaling of $z^{-2/3}$ is applied), to the sound horizon at the baryon drag epoch, $r_{d}$, shown in the upper panels, while the lower panels show the ratio of transverse and line-of-sight comoving distances $D_{M}/D_{H}$. The data includes contributions from all tracers and redshift bins, as labeled. The solid lines represent predictions from different cosmological models: BDE-CDM, $\Lambda$CDM, and the $w_0w_a$CDM models. These predictions are based on their best fits. This analysis combines data from DESI, CMB, and each of the SNe-Ia datasets. The bottom panels in each plot show how the best fits for BDE-CDM and $w_0w_a$CDM models compare to $\Lambda$CDM model.  }
\end{figure*}
\begin{figure*}
    \centering
    \includegraphics[width=\textwidth]{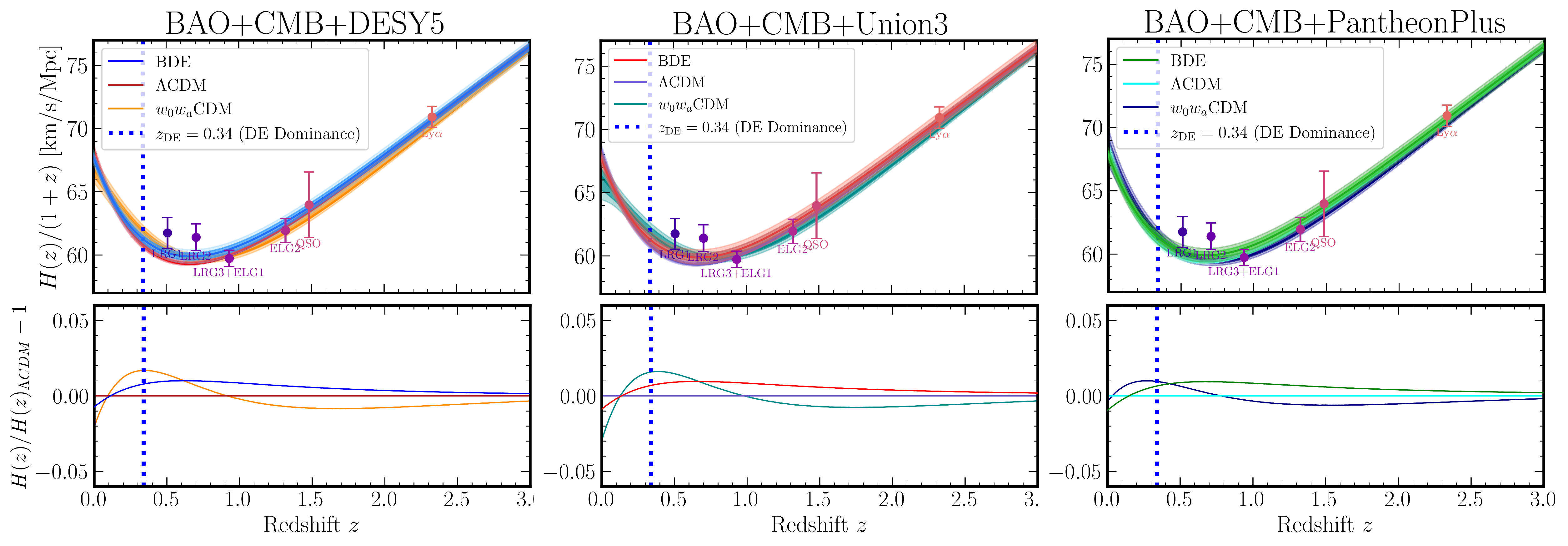}
    \caption{The comoving Hubble parameter is presented as a function of redshift, derived from DESI DR2 BAO data combined with the CMB and the PantheonPlus, Union3, and DESY5 SNe-Ia datasets. The color bands indicate the $68\%$ and $95\%$ confidence intervals permitted by the data combinations in the BDE-CDM, $w_0w_a$CDM, and $\Lambda$CDM models. These results demonstrate the onset of acceleration at approximately $z\simeq0.6$ for $\Lambda$CDM, $z\simeq0.62$ for BDE-CDM, and $z\simeq0.81$ for $w_0w_a$CDM. Colored data points represent the BAO measurements from DESI DR2. The bottom panels in each plot illustrate the fractional difference between the best-fit BDE and $w_0w_a$CDM models relative to $\Lambda$CDM. }
    \label{Fig_9:Comoving Hubble parameter}
\end{figure*}
\begin{figure*}[hb]
    \centering
    \includegraphics[width=\textwidth]{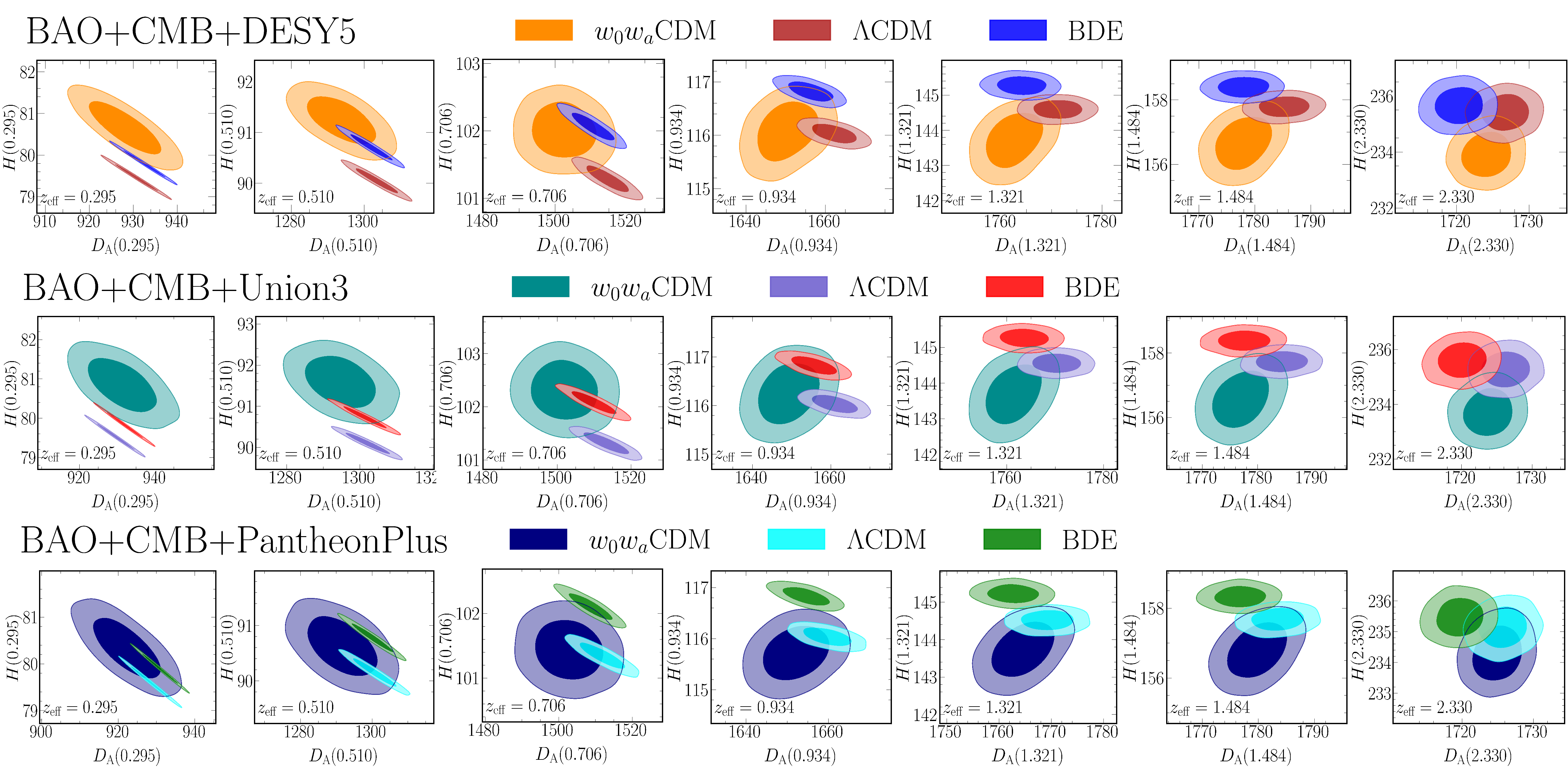}
    \caption{This figure presents constraints on the angular diameter distance $D_{A}(z)$ and the Hubble parameter $H(z)$, derived from DESI BAO + CMB + DESY5, DESI BAO + CMB + Union3 and DESI BAO + CMB + PantheonPlus at the seven central redshifts analyzed by the DESI collaboration \cite{tr6ykpc6} in DESI DR2. The plots display the $68\%$ and $95\%$ confidence contours for $D_{A}(z)$ and $H(z)$ for the BDE-CDM, $w_0w_a$CDM, and $\Lambda$CDM models. As the effective redshift increases from low to high values, the contour direction for the BDE-CDM and $\Lambda$CDM models shifts counterclockwise, whereas the $w_0w_a$CDM contours remain unchanged.}
    \label{Fig_10:Angular Diameter distance - Hubble function planes}
\end{figure*}
\begin{figure*}
    \centering
    \includegraphics[width=\textwidth]{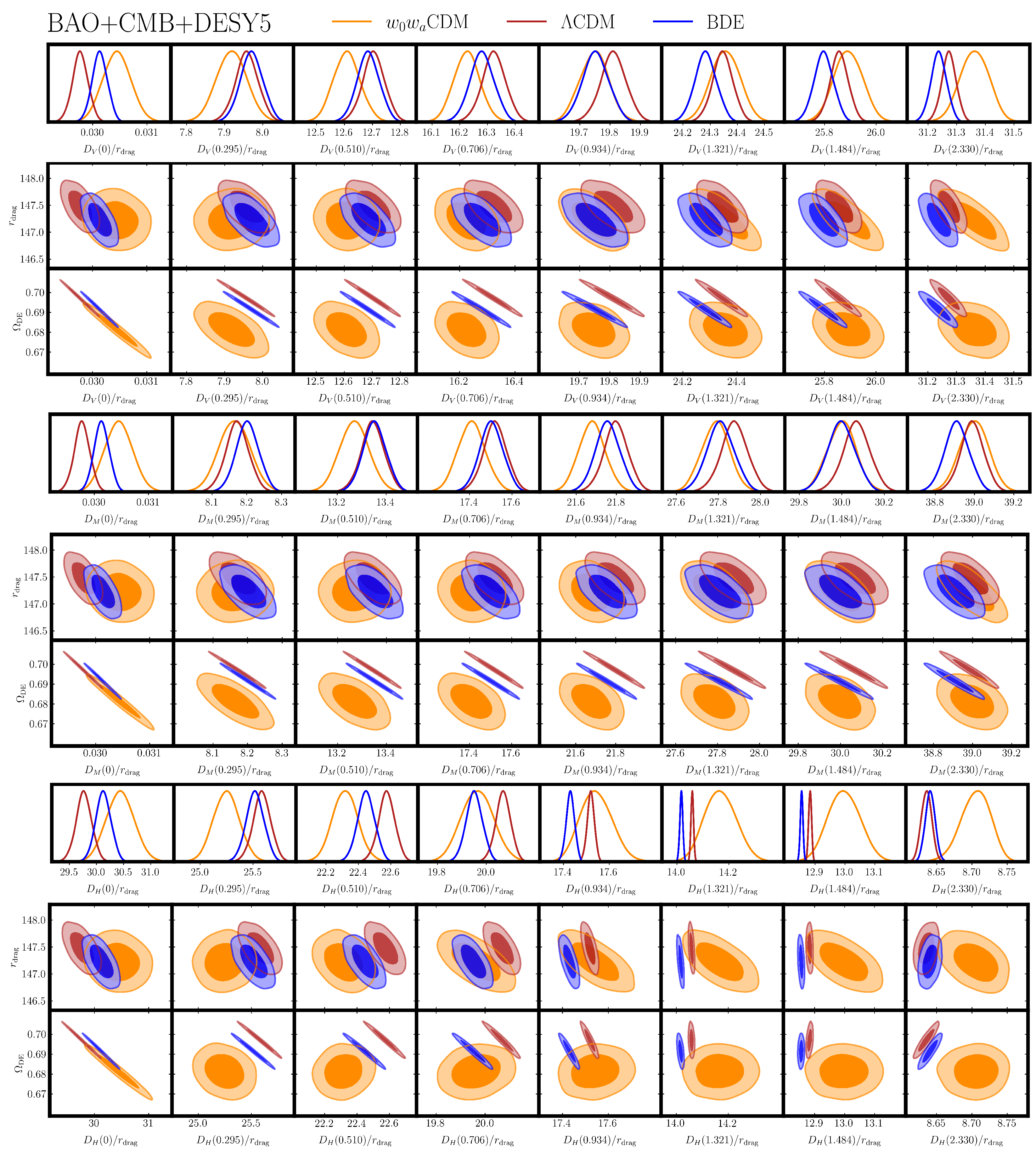}
    \caption{The upper, middle, and lower panels present the marginalized 1D posteriors and $68\%$ and $95\%$ confidence contours for the angle-average distance, which is defined as $D_{V}\equiv(zD^{2}_{M}D_{H})^{1/3}$, the comoving angular distance $D_{M}(z)$ and the Hubble distance $D_{H}(z)$ to the sound horizon at the baryon drag epoch, $r_{d}$, with respect to Hubble parameter and dark energy density parameter $\Omega_{\mathrm{DE}}$ at present time at the seven central redshifts of DESI collaboration analysis of DESI DR2 data. The marginalized 1D posteriors and confidence contours are derived from the combination of DESI BAO DR2 + CMB + DESY5 datasets in the framework of BDE, $w_0w_a$CDM and $\Lambda$CDM models. }
    \label{Fig_11:Dvrd_DMrd_DHrd_contours_rd_OmegaDE}
\end{figure*}

\subsection{Distance-redshift and expansion rate measurements}

BAO measurements span a wide range of redshifts, enabling constraints on key cosmological parameters governing the distance-redshift relation—including spatial curvature, dark energy properties, and the Hubble constant. These constraints are further strengthened when external priors on the absolute BAO scale are incorporated. The distance-redshift results from a joint analysis of DESI BAO, CMB and individual SNe-Ia datasets (PantheonPlus, Union3, and DESY5) are displayed in upper and lower panels given in Fig.~\ref{Fig_8: Distance redshift measurements} in a Hubble diagram and presents a comprehensive comparison of the measured BAO distance scales at multiple redshifts, as obtained by the DESI project \cite{tr6ykpc6}. In the upper panels of Fig.~\ref{Fig_8: Distance redshift measurements}, the solid lines represent the best-fit of each model for the angle-averaged distance, defined as $D_{V} \equiv (zD^{2}_M D_H)^{1/3}$, divided by the sound horizon at the baryon drag epoch, $r_{d}$, with an arbitrary scaling of $z^{-2/3}$ applied for visualization clarity. The DESI BAO measurements are shown across a range of redshift bins and include data from all tracers utilized in the DESI survey, as indicated by the data points with their error bars in each plot. The lower panels display the ratio of the transverse comoving distance to the line-of-sight comoving distance, $D_{M}/D_{H}$, which provides additional geometric information about the universe's expansion history. Superimposed on the data points are solid lines indicating the predicted values from three cosmological models—BDE-CDM, $\Lambda$CDM, and $w_0w_a$CDM—where each curve corresponds to the best-fit parameters derived from a joint analysis of DESI DR2 BAO, CMB, and individual SNe-Ia datasets (PantheonPlus, Union3, and DESY5). Fig.~\ref{Fig_8: Distance redshift measurements} thus enables direct visual assessment of how well each model matches the observed BAO measurements across redshifts. The bottom sub-panels in each plot from Fig.~\ref{Fig_8: Distance redshift measurements} further highlight the fractional differences between the BDE-CDM and $w_0w_a$CDM models relative to the $\Lambda$CDM prediction, emphasizing any deviations or concordances in the context of the standard cosmological model. This detailed breakdown provides insight into the sensitivity of BAO measurements to different theoretical frameworks and the constraining range of the combined datasets on cosmological parameters. Figure \ref{Fig_9:Comoving Hubble parameter} displays the comoving Hubble parameter as a function of redshift in BDE-CDM, $w_0w_a$CDM, and $\Lambda$CDM,  illustrating how well the BDE-CDM model fits the DESI DR2 BAO measurements of $H(z)$ over the redshift range $0<z<3$, with the shaded regions indicating the $68\%$ and $95\%$ credible regions obtained from our MCMC samples, yet fails to match the distance-ladder estimate of ($H_0=73.04 \pm 1.04$) $\mathrm{km/s/Mpc}$ from the SH0ES project \cite{Riess_2022} and other $H_{0}$ studies \cite{Riess_2023}. The sub-lower panels in each plot display the difference between the best-fit BDE-CDM and $w_0w_a$CDM models relative to the $\Lambda$CDM model. Observational data from DESI surveys, including LRGs, ELGs, and Ly$\alpha$-QSO, are presented as colored data points with associated error bars in Figure \ref{Fig_9:Comoving Hubble parameter}. The vertical dashed line denotes the redshift $z_{\mathrm{DE}}=0.34$, which marks the onset of dark energy dominance. Since $H(z)/(1+z)$ is directly proportional to $\dot{a}$, the expansion rate, its slope with respect to redshift corresponds to the negative of the cosmic expansion acceleration $-\ddot{a}$. The observed variation in $H(z)$ indicates that all three cosmological models require a period of accelerated expansion, represented by a negative slope, at $z\lesssim 0.8$, where it can also be seen in Fig.~\ref{fig_4: diagnostic and deceleration} in terms of $q(z)$ indicating that the cosmic acceleration$(q<0)$--according to $w_0w_a$CDM model--started to take off earlier at $z\sim0.8$.
Figure \ref{Fig_10:Angular Diameter distance - Hubble function planes} presents the $68\%$ and $95\%$ marginalized posterior constraints on the Hubble parameter $H(z)$ as a function of the comoving angular distance $D_{M}(z)$, evaluated at seven central redshifts (effective redshifts, $z_{\mathrm{eff}}$) as determined by the DESI collaboration's second data release \cite{tr6ykpc6}. These constraints provide a comprehensive overview of the expansion history of the universe as inferred from large-scale structure observations. Notably, in the low-redshift regime at approximately $z\sim0.3$ (as depicted in Figure \ref{Fig_10:Angular Diameter distance - Hubble function planes}), the $w_0w_a$CDM model demonstrates a conformal expansion rate that exceeds those predicted by the $\Lambda$CDM and BDE-CDM models. This phenomenon is attributable to the EoS parameter of the $w_0w_a$CDM model, which begins to deviate from the cosmological constant value ($w=-1$) at redshifts $z\le0.5$, reflecting an increased contribution of DE that induces a more rapid cosmological expansion during this epoch. In contrast, the BDE-CDM model predicts an expansion rate that closely tracks that of the $\Lambda$CDM model at higher redshifts ($z\ge1$), with only a minor deviation—approximately $0.8\%$—observed at $z\sim0.6$. These detailed comparative findings underscore the sensitivity of cosmological observables to the underlying DE model and highlight the potential of precision measurements in discriminating between competing theoretical frameworks. Furthermore, Fig. \ref{Fig_11:Dvrd_DMrd_DHrd_contours_rd_OmegaDE} presents the $68\%$ and $95\%$ confidence levels for the cosmological distance measures $D_{V}/r_{d}$, $D_{M}/r_{d}$, and $D_{H}(z)/r_{d}$, analyzed as functions of sound horizon at the baryon drag epoch $r_{\mathrm{drag}}$ and the DE density parameter $\Omega_{\mathrm{DE}}$. This comprehensive analysis encompasses seven redshift bins, each reflecting the BAO signals detected by the DESI collaboration \cite{tr6ykpc6} over the broad redshift interval $0.1 < z < 4.2$. The effective redshifts, $z_{\mathrm{eff}}$, are determined from precise fits to the clustering properties of DESI DR2 galaxies, quasars, and the Ly$\alpha$ forest, and are given by the sequence: 0.295, 0.510, 0.706, 0.934, 1.321, 1.484, and 2.330. Of particular interest is the observed behavior of the joint constraints in the $r_{\mathrm{drag}}$–$\Omega_{\mathrm{DE}}$ parameter space: as the effective redshift increases, the contours representing the BDE-CDM and $\Lambda$CDM models exhibit a progressive divergence across all distance measures. This separation becomes more pronounced at higher redshifts, potentially reflecting the differing underlying physics encoded within each model. Nevertheless, for the majority of the redshift bins and distance measures considered—specifically $D_{V}/r_{d}$, $D_{M}/r_{d}$, and to a lesser extent $D_{H}/r_{d}$—the contours for both BDE-CDM and $\Lambda$CDM remain statistically consistent with those of the $w_0w_a$CDM model at the $1\sigma$ confidence level. These results highlight both the discriminating power and the current limitations of state-of-the-art BAO measurements in constraining the fundamental parameters of DE models within the context of precision cosmology.

\begin{figure*}[ht]
      \includegraphics[width=\textwidth]{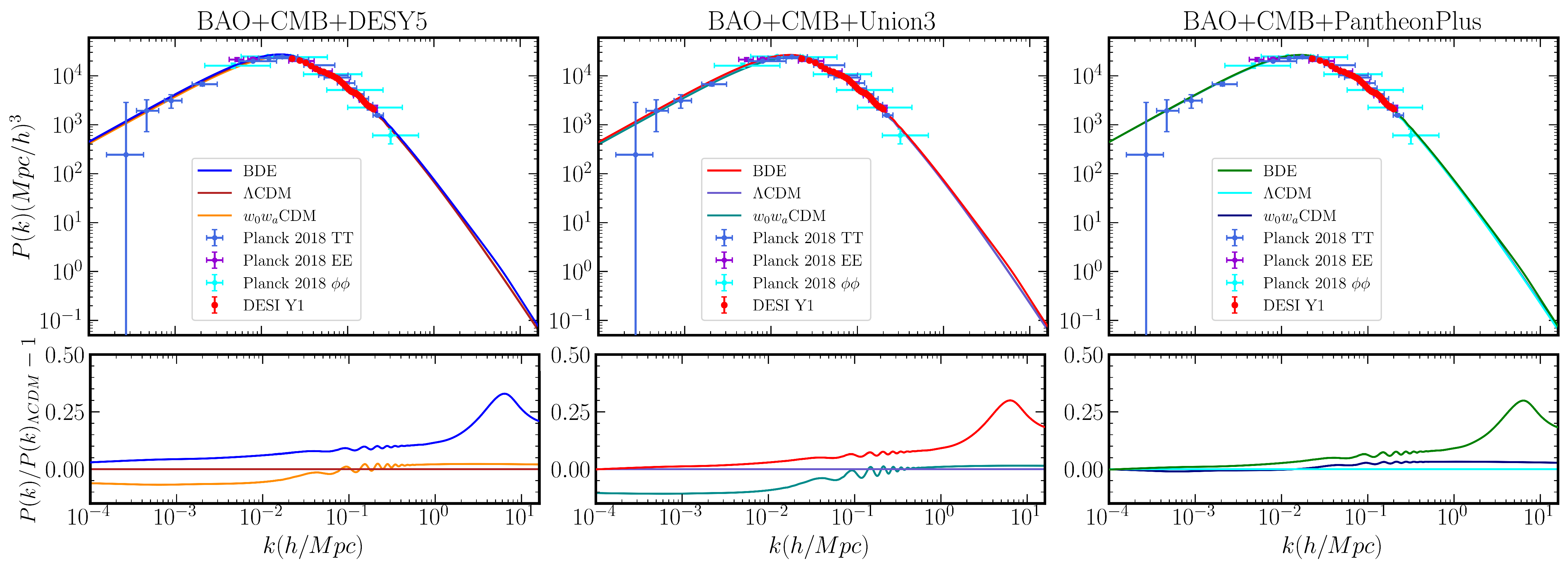}
  \caption{\label{Fig_12: matter power spectrum} The linear matter power spectrum $P(k)$ (top) at redshift $z=0$ and its fractional deviation from $\Lambda$CDM, $P(k)/P(k)_{\Lambda}CDM-1$ (bottom), for the BDE-CDM, $w_0w_a$CDM, and $\Lambda$CDM models from DESI BAO DR2 combined with CMB and each of the PantheonPlus, Union3 and DESY5 SNe-Ia datasets. All models show consistency with \textit{Planck} 2018 CMB data (TT, EE, $\phi \phi$) at all scales. BDE-CDM model exhibits a progressive enhancement of power (up to $ \sim 25\%$) at small scales ($k \gtrsim 1$), aligning with DESI DR2 BAO measurements. The $w_0w_a$CDM model also displays deviations indicating parametric flexibility in its DE-EoS.  }
\end{figure*}
\begin{figure*}
      \includegraphics[width=\textwidth]{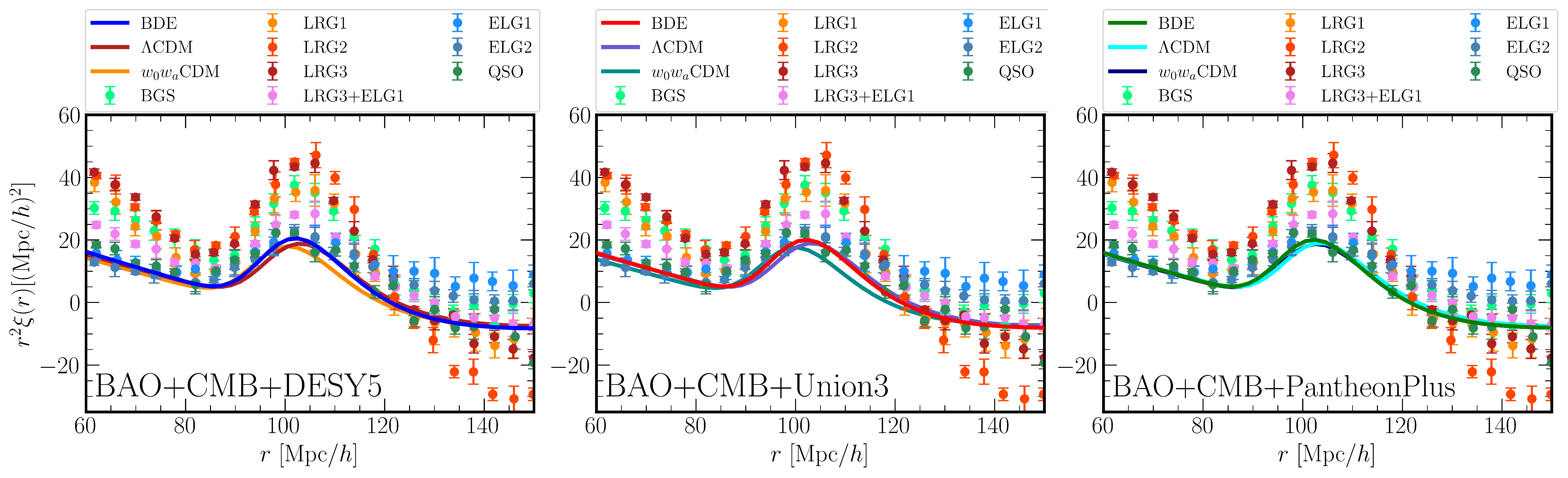}
  \caption{\label{Fig_13: Matter correlation} Correlation function of the matter power spectrum in the framework of BDE-CDM, $w_0w_a$CDM and $\Lambda$CDM models represented as colored solid lines. The multipole moments of the DESI DR2 correlation functions of galaxies and quasars, where the three panels display the monopole moments, are represented with filled circles that correspond to the data measurements, and the error bars represent the $68\%$ confidence intervals. }
\end{figure*}
\begin{figure}
  \centering
      \includegraphics[width=0.6\textwidth]{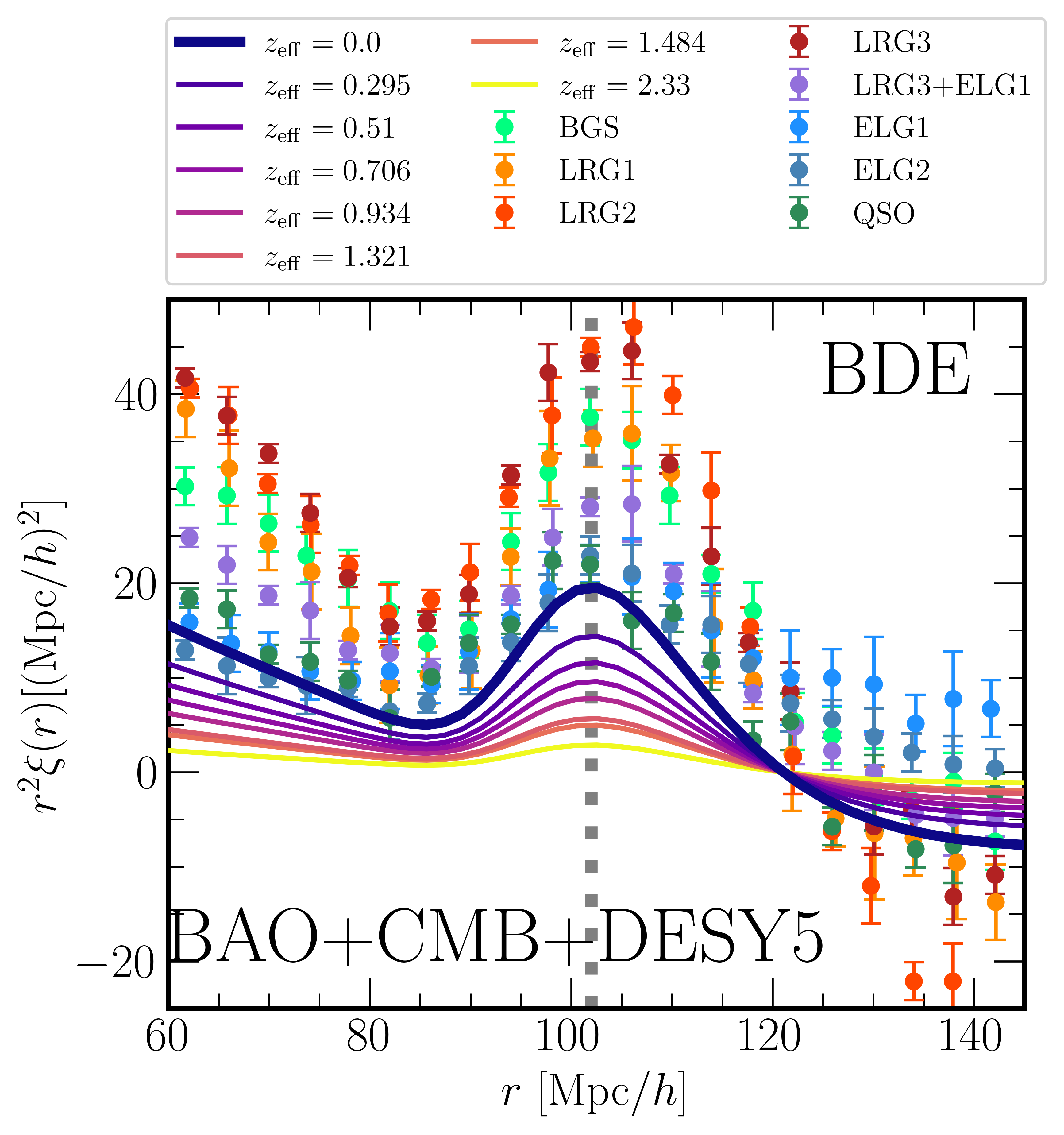}
  \caption{\label{Fig_14: evolution of the correlation function in BDE} Evolution of the baryon acoustic oscillation (BAO) peak given by the two-point correlation function at a distance corresponding to the sound horizon $r_d$ at several redshifts in the framework of the BDE-CDM model. The dotted line marks the location of the BAO peak  $\sim 102 h^{-1}$ Mpc.   }
\end{figure}
\begin{figure*}
  \centering
      \includegraphics[width=\textwidth]{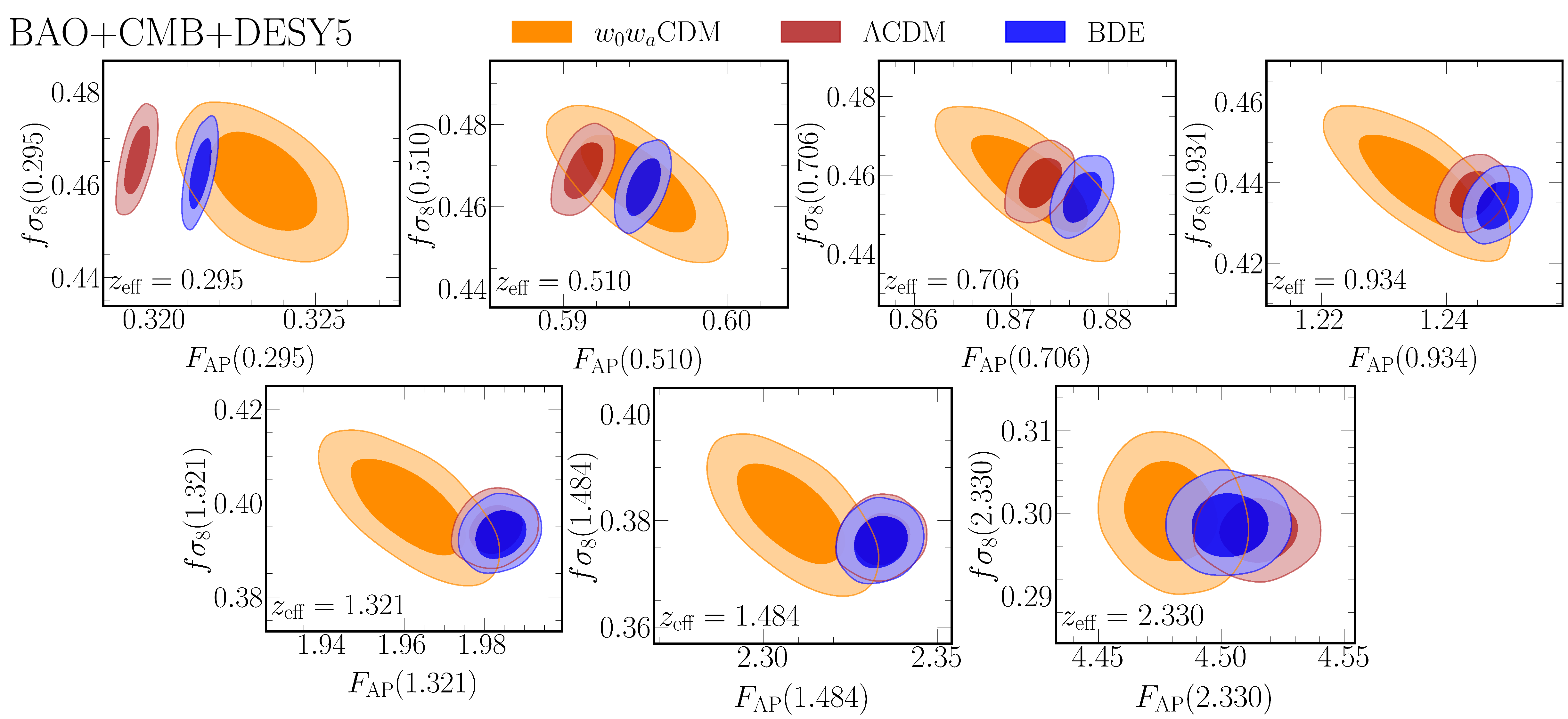}
  \caption{\label{Fig_15: fsigma and Alcock paczinski contours planes} Constraints on $f\sigma_8$ and $F_{AP}$ from analysis of redshift-space distortions. The contours show the $68\%$ and $95\%$ confidence ranges on $(f\sigma_8, F_{AP})$ from BDE-CDM (in blue), $w_0w_a$CDM (in orange), and $\Lambda$CDM (in red) at different effective redshifts.   }
\end{figure*}
\begin{figure*}
  \centering
  \begin{tabular}{c@{\hspace{1em}}c@{\hspace{1em}}c@{\hspace{1em}}c}
      \includegraphics[width=\textwidth]{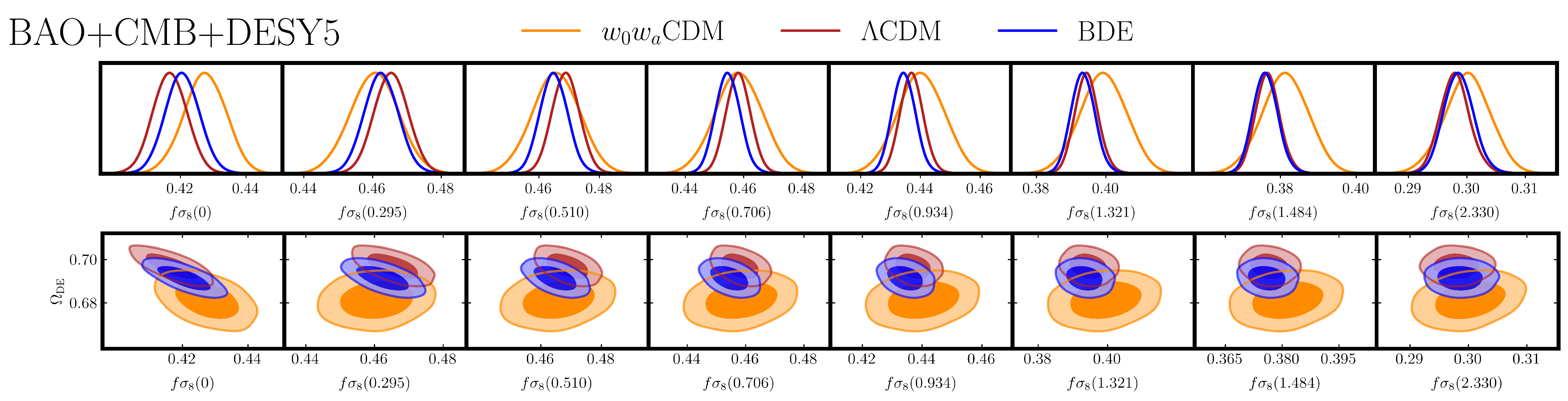}&\\
      \includegraphics[width=\textwidth]{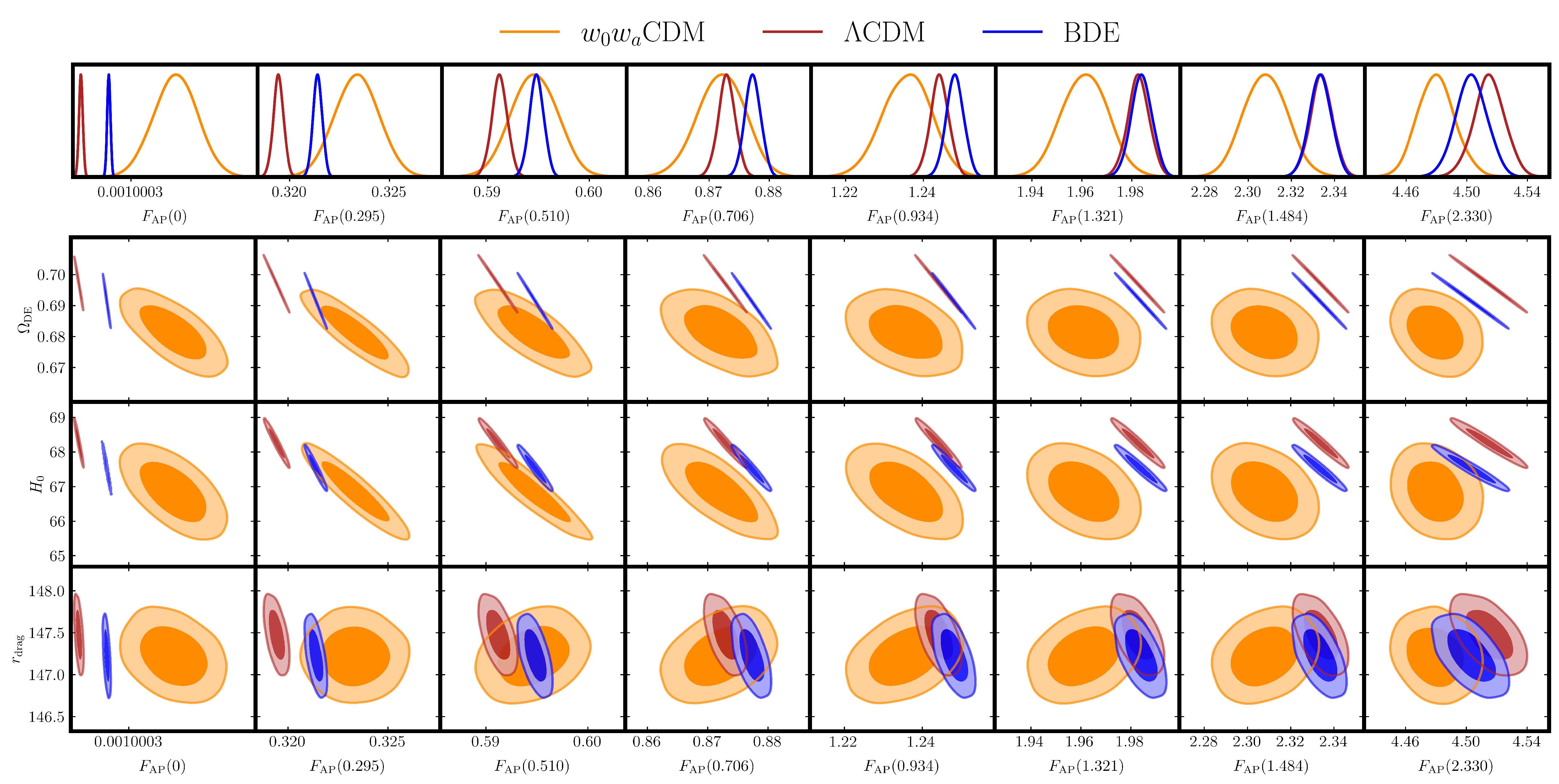} 
  \end{tabular}
  \vspace*{8pt}
  \caption{\label{Fig_16: fsigma and FAP as a function of dark energy} \textit{Upper plot}: Marginalized distributions in BDE-CDM (blue), $w_0w_a$CDM (dark orange), and $\Lambda$CDM (dark red)  of the combination $f\sigma_{8}$ at different effective redshifts where BAO signals are detected by DESI. It is displayed the joint constraints on $\Omega_{DE}$ and $f\sigma_8$ for each case are displayed. \textit{lower plot}: Marginalized distributions of the combination of Alcock-Paczinski parameter at different effective redshifts of DESI BAO signals. It is also displayed the joint constraints on $\Omega_{DE}$, $H_0$, and $D_{M}$ with $F_{AP}$ for each case.  }
\end{figure*}
\begin{figure}
  \centering
      \includegraphics[width=0.6\textwidth]{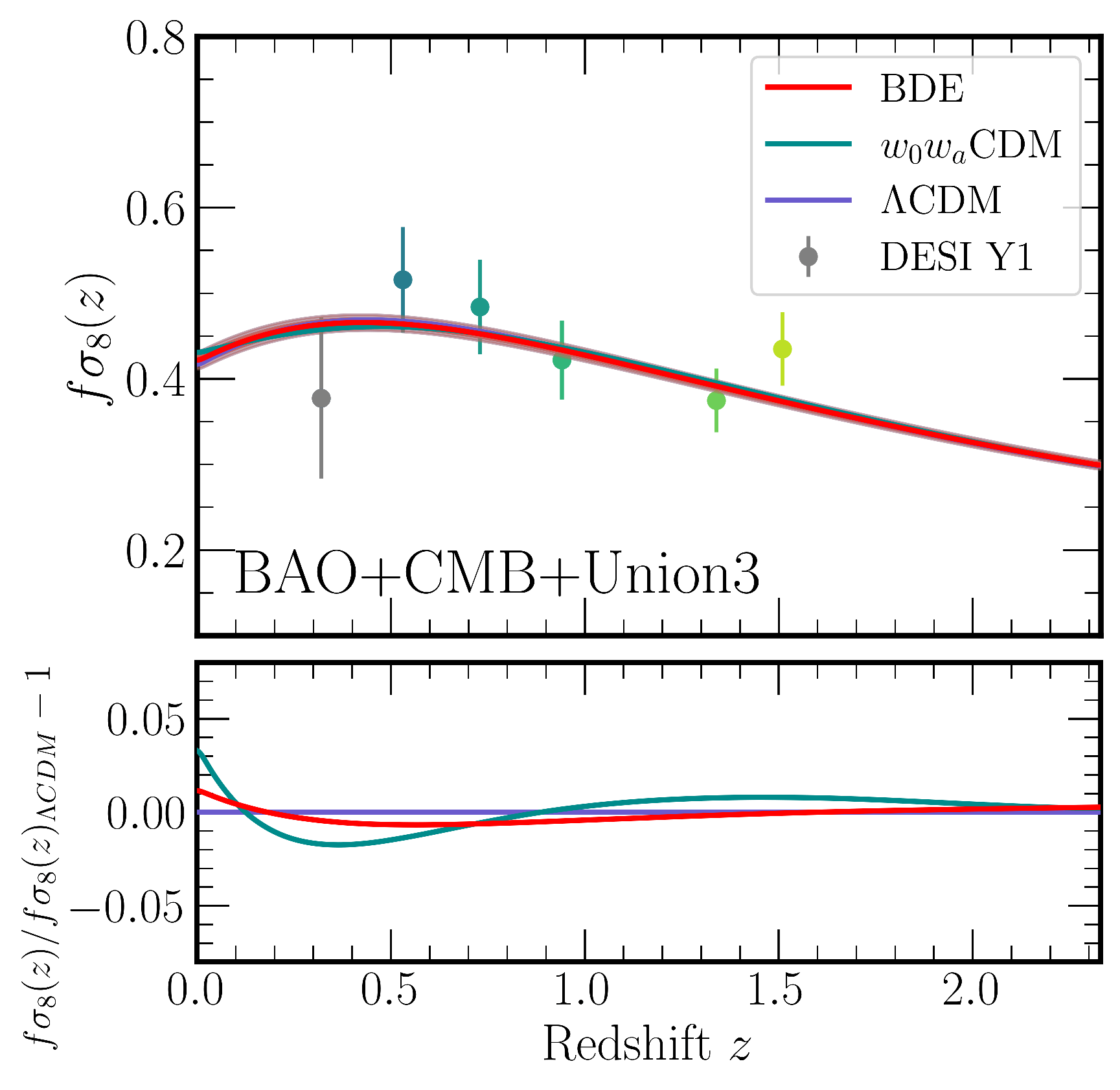}
  \caption{\label{Fig_17_ fsgima8 evolution} The growth of structure measurement parameter as a function of redshift, $f\sigma_{8}(z)$, for BDE-CDM, $w_0w_a$CDM, and $\Lambda$CDM plotted along with the shaded bands correspond to $68\%$ and $95\%$ confidence regions for each model from DESI BAO combined with CMB and Union3. In the bottom panel, it is shown the relative difference of BDE-CDM and $w_0w_a$CDM with respect to $\Lambda$CDM.}
\end{figure}
\subsection{Matter power spectrum and structure formation}

 It is presented in Fig.~\ref{Fig_12: matter power spectrum} the linear matter power spectrum at the present epoch ($z=0$) for BDE-CDM, $w_0w_a$CDM, and $\Lambda$CDM models, derived from DESI BAO DR2 in combination with CMB and each SNe-Ia compilation dataset. The lower panels display the fractional deviations of BDE-CDM and $w_0w_a$CDM from the baseline $\Lambda$CDM model, alongside observational data from the \textit{Planck} 2018 mission \cite{refId0} and DESI Year 1 survey \cite{cereskaite2025inferencematterpowerspectrum}, establishing a consistent framework for evaluating cosmic structure evolution across redshifts. The power spectra for each model are computed within the range $0.0001\,h\, \mathrm{Mpc}^{-1} \leq k \leq 10\,h\, \mathrm{Mpc}^{-1}$, as shown in Fig.\ref{Fig_12: matter power spectrum}. Examining the characteristics of $P(k)$ for BDE-CDM relative to $\Lambda$CDM reveals notable differences. At large scales ($k < 0.01$ $h\, \mathrm{Mpc}^{-1}$), only minor deviations in $\Delta P(k)/P(k)_{\Lambda\mathrm{CDM}}$ are observed, attributable to the distinct primordial power spectrum of the BDE-CDM model. This result is anticipated, as these long-wavelength modes cross the horizon at later epochs and thus have less time to evolve.
At intermediate and small scales ($k \geq 0.05\, h\, \mathrm{Mpc}^{-1}$), a marked increase in the growth of density fluctuations is observed, with $\Delta P(k) > 0$. Initially, the amplitude of modes crossing the horizon before $a_c$ is suppressed relative to $\Lambda$CDM and $w_0w_a$CDM, due to free-streaming particles from the DGG group ($a < a_c$). Subsequently, these modes are enhanced as a result of the rapid dilution of the energy density of our Bound Dark Energy model with $ \rho_{\rm BDE}\sim a^{-6}, w_{\rm DE}\sim 1$ following the condensation scale transition \cite{PhysRevLett.121.161303,PhysRevD.72.043508}. Consequently, the matter power spectrum in the BDE-CDM model is enhanced at all small scales compared to $\Lambda$CDM and $w_0w_a$CDM, with an increase of approximately $25\%$ at the peak of the bump near $k \approx 4.3\, \mathrm{Mpc}^{-1}$ \cite{PhysRevLett.121.161303,PhysRevD.72.043508}. This enhancement reflects the unique expansion history and DE dynamics of the BDE-CDM model, demonstrating its impact on cosmic structure growth. In contrast, the $w_0w_a$CDM model exhibits a slight net positive deviation from $\Lambda$CDM at small scales. At even smaller scales ($k > 1\, h\, \mathrm{Mpc}^{-1}$), deviations in the matter power spectrum stabilize as nonlinear effects become dominant; a detailed analysis of these effects is reserved for future work. This comparison highlights the complexity and diversity of cosmic structure formation under different DE scenarios.

 \noindent We will now expand our analysis of the matter power spectrum by examining its Fourier transform: the two-point correlation function (2PCF), $ \xi(r) $. The function $ \xi(r) $ serves as a vital statistical metric for large-scale structure (LSS), encapsulating information about matter clustering and its evolution due to gravitational collapse. A key feature of $ \xi(r) $ is the BAO peak, which is a remnant of sound waves propagating through the primordial plasma. This peak appears as a localized enhancement in clustering at the sound horizon scale, approximately $ r_{d} \sim 100 h^{-1} \mathrm{Mpc} $. This characteristic has proven to be a reliable standard ruler for constraining cosmological models. Figure \ref{Fig_13: Matter correlation} presents the best-fit function $\xi(r)$ from BDE-CDM, $w_0w_a$CDM, and $\Lambda$CDM models, alongside observational data derived from DESI tracers, in which the acoustic feature is clearly noticeable as a distinct peak in the correlation around $100h^{-1} \text{ } \mathrm{Mpc}$ and for all DESI tracers \cite{tr6ykpc6}. The observational DESI tracers, including BGS, LRGs, ELGs, and QSOs, present deviations largely attributed to variations in galaxy bias, redshift distributions, and survey systematics. The BAO peak position remains firmly anchored to the sound horizon $r_d$, governed by pre-recombination physics. For the BDE-CDM model, the sound horizon aligns with that of $\Lambda$CDM when calibrated against CMB data, thereby ensuring that the BAO peak's position is consistent with observational evidence. This consistency arises because $r_d$ is predominantly determined by early-universe physics, which remains unaffected by the late-time dynamics inherent to BDE. A detailed examination of Figure \ref{Fig_13: Matter correlation} reveals that the amplitude BAO peak in the BDE-CDM model exhibits a modest but discernible increase relative to predictions from the standard $\Lambda$CDM scenario. This behavior is fundamentally attributable to the distinct evolution of the DE-EoS within the BDE-CDM framework. As previously discussed, the EoS for the BDE-CDM model transitions from a stiff phase ($w_{\mathrm{BDE}}=1$) during the early universe, progressing towards an epoch of cosmological constant-like behavior ($w\sim-1$), and ultimately ending up at $w_{\mathrm{BDE}} \approx -0.93$ at the present epoch. This implies a rapid dilution that temporarily lowers the DE density, consequently affecting the structure growth rate and the clustering amplitude at late times. Furthermore, the inclusion of DGG relativistic particles in the BDE-CDM model serves to elevate the total radiation energy density, thereby modifying the sound speed of the early baryon-photon plasma. This alteration enhances the amplitude of BAO, manifesting as positive deviations in the two-point correlation function at separations $r \sim r_d$, where $r_d$ denotes the sound horizon at the drag epoch. Consequently, the BDE-CDM model yields a correlation function that consistently exceeds that of $\Lambda$CDM at the BAO scale, unambiguously reflecting the augmented peak amplitude and highlighting the interplay between evolving DE physics and large-scale structure observables as seen in Fig.\ref{Fig_13: Matter correlation}. Figure \ref{Fig_14: evolution of the correlation function in BDE} illustrates the evolution of the BAO peak in the BDE-CDM model across different redshifts, in which the DESI collaboration measured the BAO signal from different tracers. In Fig.~\ref{Fig_15: fsigma and Alcock paczinski contours planes}, we present a marginalized contours for the $ f\sigma_{8}$ (with $f$ representing the growth rate of the matter density perturbations and $\sigma_{8}$ is the matter variance on an $8h^{-1}$ Mpc scale) and the Alcock-Paczynski $F_{AP}$ parameters, across seven distinct redshift bins analyzed by DESI collaboration \cite{tr6ykpc6}. The $F_{AP}$ parameter is formulated as $F_{AP}(z) = D_M(H(z)/c) $, where $ D_M$ is the comoving angular diameter distance, $H(z)$ is the Hubble parameter at redshift $z$, and $c$ denotes the speed of light. This parameter is essential for correcting any geometric distortions due to the expansion of the universe, enabling accurate comparisons between the observed and theoretical distributions of galaxies. The upper plot in Fig.~\ref{Fig_16: fsigma and FAP as a function of dark energy} displays the joint constraints of $f\sigma_{8}$ with respect to $\Omega_{DE}$ at the same effective redshifts as shown in Fig.~\ref{Fig_15: fsigma and Alcock paczinski contours planes}. While the confidence regions align within the $1\sigma$ level, we note that the relation follows: $f\sigma_8(\mathrm{BDE\text{-}CDM}) < f\sigma_8(\Lambda\mathrm{CDM}) < f\sigma_8(w_0w_a\mathrm{CDM})$. However, the observed differences among these models are quite subtle, indicating no significant tension present in the results. The lower plot in Fig.~\ref{Fig_16: fsigma and FAP as a function of dark energy} displays the joint constraints of the Alcock-Paczynski $F_{AP}$ with respect to $\Omega_{DE}$, $H_0$, and $r_{\mathrm{drag}}$. In Fig.~\ref{Fig_17_ fsgima8 evolution}, we present the best-fit values along with the $68\%$ and $95\%$ confidence bands for the evolution of the structure growth parameter, $f\sigma_{8}(z)$ in the BDE-CDM, the $w_0w_a$CDM, and $\Lambda$CDM models from DESI+CMB+Union3 datasets. The confidence bands illustrate the uncertainties associated with the growth measurements as a function of redshift. Additionally, we include the observational results from the six distinct redshift bins derived from DESI DR1 data, which are indicated by filled circles, and the error bars represent the $68\%$ confidence intervals. These measurements provide critical insights into cosmic structure formation and help constrain the parameters governing dark energy evolution and the expansion history of the universe.
\begin{table}[ht]
\centering
    \begin{tabular}{lcccc}
    \hline
    \textrm{Model/Dataset} & \multicolumn{2}{c}{ $\mathbf{BDE}$-\textbf{CDM}}  & \multicolumn{2}{c}{ $\mathbf{w_0w_a}$\textbf{CDM}}  \\ 
    \hline 
                     & $\Delta$DIC  & $\Delta$AIC  &  $\Delta$DIC & $\Delta$AIC    \\ 
    \hline
    BAO              & -2.85  & -4.89  & -1.18  & -0.85   \\ 
    PantheonPlus     & -3.23  & -5.23  & -0.62  &  1.27   \\
    Union3           &  1.10  & -1.05  & -1.83  & -1.21   \\ 
    DESY5            & -0.32  & -2.30  & -0.82  & -2.50   \\ 
    BAO+PantheonPlus & -3.72  & -5.80  & -1.83  & -1.21   \\ 
    BAO+Union3       & -4.21  & -6.57  & -3.37  & -3.14  \\ 
    BAO+DESY5        & -6.77  & -8.97  & -4.27  & -4.25   \\ 
    \hline
	\end{tabular}
    \caption{\label{tab:cosmological_fit} Information criteria comparison of cosmological models under study. $\Delta\mathrm{AIC}$ and $\Delta\mathrm{DIC}$ values relative to $\Lambda$CDM (reference model) for BDE-CDM and $w_0w_a$CDM models across BAO, SNe-Ia (PantheonPlus, Union3, DESY5), and combined datasets. Models contain $k = 5$, $6$, and $8$ free parameters for BDE-CDM, $\Lambda$CDM, and $w_0w_a$CDM, respectively. Negative values favor the alternative model.} 
\end{table}
\section{Model comparison}

To quantify whether the inclusion of dark energy dynamics improves the cosmological fit to late-time observations compared to the $\Lambda$CDM baseline, it requires the application of robust statistical criteria that balance goodness-of-fit against model complexity. To this end, we employ two information criteria: the Deviance Information Criterion (DIC) \cite{Spiegelhalter, Liddle} and the Akaike Information Criterion (AIC) \cite{akaike1974new}. The DIC is defined as $\text{DIC} \equiv \overline{D(\theta)} + 2p_D$, where the deviance $D(\theta) = -2 \ln \mathcal{L}(\theta)$ quantifies the log-likelihood of the data given the model parameters $\theta$, $\overline{D(\theta)}$ represents the posterior mean of the deviance averaged over the parameter space, and the effective number of parameters $p_D = \overline{D(\theta)} - D(\bar{\theta})$ serves as a complexity penalty that accounts for the model's flexibility. The AIC is given by $\text{AIC} \equiv 2k - 2 \ln \mathcal{L}_{\text{max}}$, where $k$ denotes the number of free parameters and $\mathcal{L}_{\text{max}}$ is the maximum likelihood value achieved by the model. Both criteria penalize unnecessary complexity while rewarding improved fits to observational data, thereby providing a principled framework for model selection that guards against overfitting. For comparative purposes, we compute the relative differences $\Delta\text{DIC} \equiv \text{DIC}_{\text{model}} - \text{DIC}_{\Lambda\text{CDM}}$ and $\Delta\text{AIC} \equiv \text{AIC}_{\text{model}} - \text{AIC}_{\Lambda\text{CDM}}$, adopting the standard $\Lambda$CDM cosmology as the reference model. Negative values indicate that the alternative model provides a superior description of the data relative to $\Lambda$CDM model. Following the conventional interpretation scale established in the literature \cite{10.1093/mnras/stw2028}, we classify the strength of evidence such that $|\Delta| < 2$ represents inconclusive evidence, $2 \leq |\Delta| < 5$ indicates positive preference, $5 \leq |\Delta| < 10$ constitutes strong evidence, and $|\Delta| \geq 10$ is considered decisive. These thresholds provide a standardized framework for interpreting the statistical significance of model comparisons.

The statistical comparison results, summarized in Table~\ref{tab:cosmological_fit}, reveal a consistent preference for dynamical dark energy models over the standard $\Lambda$CDM model, with BDE-CDM emerging as the statistically favored framework across most dataset combinations. When considering the DESI DR2 BAO measurements alone, the BDE-CDM model achieves $\Delta\text{DIC} = -2.85$ and $\Delta\text{AIC} = -4.89$, whose values indicate positive evidence in favor of BDE-CDM according to the Jeffreys' scale. This improvement is particularly noteworthy given that BDE-CDM employs one parameter less than $\Lambda$CDM, demonstrating that the enhanced fit quality originates from the physically motivated dynamics of the DE-EoS rather than from increased model flexibility. The $w_0w_a$CDM model also shows improvement over $\Lambda$CDM with $\Delta\text{DIC} = -1.18$ and $\Delta\text{AIC} = -0.85$; however, this preference remains in the inconclusive regime and, despite possessing three additional free parameters relative to BDE-CDM model, achieves a smaller improvement in data fit.

The analysis of Type Ia supernovae data yields complementary insights into the relative performance of each model. For the PantheonPlus compilation, BDE-CDM exhibits particularly strong performance with $\Delta\text{DIC} = -3.23$ and $\Delta\text{AIC} = -5.23$, the latter exceeding the threshold for strong evidence. The $w_0w_a$CDM model, by contrast, achieves only marginal improvement in DIC ($\Delta\text{DIC} = -0.62$) and is actually disfavored by the AIC ($\Delta\text{AIC} = +1.27$), reflecting the penalty imposed by its additional parameters without improvement in data fit. The Union3 compilation presents the most conservative results for BDE-CDM, with $\Delta\text{DIC} = +1.10$ indicating a slight preference for $\Lambda$CDM while $\Delta\text{AIC} = -1.05$ shows inconclusive preference for BDE-CDM; this apparent tension between criteria reflects the different weighting of complexity penalties in DIC versus AIC. For the DESY5 dataset, both dynamical dark energy models demonstrate comparable performance, with BDE-CDM obtaining $\Delta\text{DIC} = -0.32$ and $\Delta\text{AIC} = -2.30$, and $w_0w_a$CDM achieving $\Delta\text{DIC} = -0.82$ and $\Delta\text{AIC} = -2.50$. The AIC values for both models approach the threshold of positive evidence, suggesting that the DESY5 supernovae data independently support deviation from a cosmological constant.

The statistical discrimination between models becomes substantially more pronounced when combining BAO measurements with SNe-Ia data, as these probes sample complementary aspects of the cosmic expansion history at late times and their joint analysis breaks parameter degeneracies inherent to individual datasets. The combination of BAO with PantheonPlus yields $\Delta\text{DIC} = -3.72$ and $\Delta\text{AIC} = -5.80$ for BDE-CDM, representing strong evidence in favor of the model, while $w_0w_a$CDM achieves $\Delta\text{DIC} = -1.83$ and $\Delta\text{AIC} = -1.21$, remaining in the inconclusive regime. When BAO is combined with Union3, both models show improved performance: BDE-CDM achieves $\Delta\text{DIC} = -4.21$ and $\Delta\text{AIC} = -6.57$, with the latter clearly exceeding the threshold for strong evidence, while $w_0w_a$CDM obtains $\Delta\text{DIC} = -3.37$ and $\Delta\text{AIC} = -3.14$, indicating positive evidence. The most compelling statistical results emerge from combining DESI DR2 BAO measurements with the Dark Energy Survey Year 5 supernovae compilation. For this dataset combination, the BDE-CDM model achieves $\Delta\text{DIC} = -6.77$ and $\Delta\text{AIC} = -8.97$, both values lying firmly within the regime of strong evidence favoring BDE-CDM over the standard $\Lambda$CDM cosmology. These represent the largest improvements observed across all dataset combinations considered in this analysis. The $w_0w_a$CDM parameterization also demonstrates strong performance with this data combination, obtaining $\Delta\text{DIC} = -4.27$ and $\Delta\text{AIC} = -4.25$, which constitutes positive evidence approaching the strong evidence threshold. However, even with three additional free parameters relative to BDE-CDM, the phenomenological $w_0w_a$CDM model fails to match the statistical improvement achieved by the BDE-CDM model.


\section{Conclusions}
\label{sec:discussion}

The results presented in this work establish the Bound Dark Energy Cold Dark Matter (BDE-CDM) model as a compelling theoretical framework that bridges fundamental particle physics with cosmology. Unlike phenomenological approaches, the BDE-CDM model provides a theoretical physical mechanism rooted in non-perturbative gauge dynamics, where dark energy emerges naturally from the condensation of the lightest meson field $\phi$ within a supersymmetric $SU(N_c=3)$ dark gauge group with $N_f = 6$ flavors, analogous to hadron formation in the strong "QCD" interaction. The EoS evolution in BDE-CDM exhibits interesting transitions: relativistic behavior ($w = 1/3$) before condensation with $N_{\rm ext} = 0.945$ extra relativistic species, a kinetic-dominated stiff phase ($w \simeq 1$, $\rho_{\rm BDE} \propto a^{-6}$) that naturally conceals dark energy during radiation and matter domination, and subsequent evolution toward the present value $w_0 = -0.9298 \pm 0.0003$. Crucially, the EoS of our BDE-CDM model is $w > -1$ throughout cosmic history, avoiding the phantom regime ($w < -1$) and associated theoretical instabilities in the $w_0w_a$CDM model when constrained by current data. The contrast in the $w_0$-$w_a$ plane area is striking: BDE-CDM confidence contour is approximately 10,000 times smaller than those of the $w_0w_a$CDM model while achieving comparable fits. The reduction in the area size is due to the lack of free parameterrs in our BDE model. Furthermore, the predictions exhibit remarkable stability across supernova compilations ($w_0 = -0.9296$ to $-0.9299$ with PantheonPlus, Union3, and DESY5), contrasting with the substantial dataset-dependent shifts in $w_0w_a$CDM model. Statistical analysis provides strong evidence favoring BDE-CDM: for BAO+DESY5, the model achieves $\Delta{\rm DIC} = -6.77$ and $\Delta{\rm AIC} = -8.97$ relative to $\Lambda$CDM and $w_0w_a$CDM models, despite employing fewer parameters. The BDE-CDM model predicts a $\sim 25\%$ enhancement in the matter power spectrum at $k \approx 4.3\,{\rm Mpc}^{-1}$.

Our BDE-CDM model successfully addresses the recent DESI observed preference for a dynamical dark energy while providing a physical mechanism rooted in gauge theories, same formalism as the  standard model interactions of particle physics, allowing therefore for a unified interpretation between dark energy and the SM. Future observations from DESI \cite{DESI_Collaboration_2022}, LSST \cite{leanne_p_guy_2024_11110648}, and Euclid \cite{euclid_experiment} will enable more stringent tests of the nature and dynamics of Dark Energy, including the small-scale power spectrum enhancement and modified BAO amplitude which is sensitive to the relative contributions of baryons and cold dark matter to structure formation. Our Bound Dark Energy model establishes a unified framework connecting particle physics and cosmology, explaining dark energy as a natural consequence of non-perturbative gauge dynamics rather than an arbitrary cosmological constant, positioning it as a viable explanation to the nature of dark energy.


\acknowledgments

We acknowledge financial support from PAPIIT-DGAPA-UNAM (101124), SECIHTI project CBF-2025-G-1299, and SECIHTI scholarship grant for Jose Lozano.


\clearpage
\bibliographystyle{JHEP}


\end{document}